\documentclass[useAMS,usenatbib]{mn2e}
\input{epsf}
\usepackage{amssymb}
\usepackage{natbib}
\usepackage{color}
\usepackage{graphicx}
\usepackage{journals}
\usepackage{multirow}
\usepackage{booktabs,threeparttable}

\newbox\grsign \setbox\grsign=\hbox{$>$} \newdimen\grdimen \grdimen=\ht\grsign
\newbox\simlessbox \newbox\simgreatbox
\setbox\simgreatbox=\hbox{\raise.5ex\hbox{$>$}\llap
     {\lower.5ex\hbox{$\sim$}}}\ht1=\grdimen\dp1=0pt
\setbox\simlessbox=\hbox{\raise.5ex\hbox{$<$}\llap 
     {\lower.5ex\hbox{$\sim$}}}\ht2=\grdimen\dp2=0pt

\definecolor{darkgreen}{rgb}{0.0,0.5,0.0}
\definecolor{darkred}{rgb}{0.5,0.0,0.0}
\definecolor{brown}{rgb}{0.65,.16,0.16}
\definecolor{grey}{rgb}{0.4,0.5,0.6}
\definecolor{darkmagenta}{rgb}{0.55,0.,0.55}
\definecolor{darkorange}{rgb}{1.,0.55,0.}

\newcommand{\hMpc}{{\ifmmode{h^{-1}{\rm Mpc}}\else{$h^{-1}$Mpc }\fi}}
\newcommand{\hGpc}{{\ifmmode{h^{-1}{\rm Gpc}}\else{$h^{-1}$Gpc }\fi}}
\newcommand{\hkpc}{{\ifmmode{h^{-1}{\rm kpc}}\else{$h^{-1}$kpc }\fi}}
\newcommand{\hMsun}{{\ifmmode{h^{-1}{\rm {M_{\odot}}}}\else{$h^{-1}{\rm{M_{\odot}}}$}\fi}}
\newcommand{\Msun}{{\ifmmode{{\rm {M_{\odot}}}}\else{${\rm{M_{\odot}}}$}\fi}}

\title[Search strategies for gravitationally lensed supernovae]
{Magnified or multiply imaged? -- Search strategies for 
gravitationally lensed supernovae in wide-field surveys}

\author[Wojtak et al.]{Rados{\l}aw Wojtak$^{1}$\thanks{E-mail:
radek.wojtak@nbi.ku.dk}, Jens Hjorth$^{1}$ and Christa Gall$^{1}$\\
$^{1}$DARK, Niels Bohr Institute, University of Copenhagen, Lyngbyvej 2, 2100 Copenhagen, Denmark \\
}

\begin{document}

\maketitle

\begin{abstract}
Strongly lensed supernovae can be detected as multiply imaged or highly magnified transients. In order to compare 
the performances of these two observational strategies, we calculate expected discovery rates as a function of survey depth 
in five $grizy$ filters and for different classes of supernovae (types Ia, IIP, IIL, Ibc and IIn). We find that detections 
via magnification is the only effective strategy for relatively shallow pre-LSST surveys. For survey depths about the LSST 
capacity, both strategies yield comparable numbers of lensed supernovae. Supernova samples from the two methods are to 
a large extent independent and combining them increases detection rates by about 50 per cent. While the number of lensed 
supernovae detectable via magnification saturates at the limiting magnitudes of LSST, detection rates of multiply 
imaged supernova still go up drastically at increasing survey depth.
Comparing potential discovery spaces, we find that lensed supernovae found 
via image multiplicity exhibit longer time delays and larger image separations making them more suitable for 
cosmological constraints than their counterparts found via magnification.

We provide useful fitting functions approximating the computed discovery rates for different supernova classes 
and detection methods. We find that the Zwicky Transient Factory will find about 2 type Ia and 4 core-collapse 
lensed supernovae per year at a limiting magnitude of 20.6 in the $r$ band. Applying a hybrid method which combines searching 
for highly magnified or multiply imaged transients, we find that LSST will detect 89 type Ia and 254 core-collapse lensed 
supernovae per year. In all cases, lensed core-collapsed supernovae will be dominated by type IIn supernovae contributing 
to 80 per cent of the total counts, although this prediction relies quite strongly on the adopted spectral templates for 
this class of supernovae. 
Revisiting the case of the lensed supernova iPTF16geu, we find that it is consistent within the 
$2\sigma$ contours of predicted redshifts and magnifications for the iPTF survey.

\end{abstract}


\section{Introduction}

The phenomenon of strongly lensed (multiply imaged) supernovae 
has long been theoretically considered \citep{Ref1964}, but the first 
detections became possible only very recently. \citet{Qui2014} found a strongly lensed Type Ia supernova 
magnified by a factor of 30, although multiple images were not resolved. The first fully resolved image configuration of 
a lensed supernova was reported by \citet{Kel2015}. The supernova (SN Refsdal) was a core-collapse type \citep{Kel2016} 
and lensed by an intervening galaxy cluster and a massive galaxy in the cluster. 
The second example of a fully resolved lensed supernova was iPTF16geu 
\citep{Goo2017}. Detected as an exceptionally luminous supernova (for its redshift) in a regular transient survey, 
the intermediate Palomar Transient Factory (iPTF), it was subsequently observed by ESO VLT, Keck Observatory, and
the Hubble Space Telescope whose images revealed a quadrupole lensing configuration with a sub-arcsec scale of image 
separations. The supernova was classified as a type Ia SN \citep{Goo2017,Can2018}. 
High-cadence imaging of massive galaxy clusters with HST also recently resulted in the discovery of a new type of 
lensed transients which appeared to be strongly lensed individual stars \citep{Rod2018,Kel2018,Chen2019,Kau2019}.

Strongly lensed supernovae are unique objects in several respects, making them a promising tool for constraining 
cosmological parameters such as the Hubble constant \citep{Gri2018,Vega2018}. With well-studied light curves, 
reasonably well-represented by simple models, lensed supernovae appear to be suitable for precise measurements of 
time delays from even relatively short observing campaigns \citep{Rod2016}. 
If, in addition, a lensed supernova is of Type Ia, 
extra constraints from the standard candle nature come into play. This can be used to measure magnification in an 
independent way \citep{Rod2015} and thus provide additional constraints on the lens model. This extra information 
can potentially narrow down inherent degeneracies in lens models which leave all lensing observables invariant, 
except the time delay \citep{Sch2014}. The unresolved degeneracies are the main source of potential 
systematic errors in cosmological constraints obtained from time-delay observations \citep{Kol1998,Ogu2003}.

Strongly lensed supernovae will be found in large numbers and in a more automatic way in ongoing and future 
transient surveys \citep[see e.g.][]{Diego2018}. Two possible strategies for finding plausible candidates can be considered. One can search for 
multiply imaged 
supernovae \citep{Ogu2010a} or highly magnified (unresolved) transients \citep{Gold2017}. The former probes a unique feature 
indicating unambiguously the lensing nature of a candidate. The latter approach is less direct and involves 
an estimate of how much brighter an observed supernova is compared to a fiducial reference supernova, as it would
have been observed in the lens galaxy (or the apparent host galaxy). Atypically bright supernovae in this case would
indicate a high chance of observing a higher-redshift, gravitationally magnified (amplified) supernova.

Searching for lensed supernovae via image multiplicity or magnification are the main observational strategies 
based on two characteristic features of the lensing phenomenon. The expected discovery rates have been estimated 
in several studies; however, each of them considering only one of the two methods and adopting specifications 
of upcoming surveys \citep{Ogu2010a,Gold2017,Gold2018}. At present, it is unclear to what extent 
the two methods are equivalent or complementary, which technique is more effective in finding lensed supernovae 
in different ranges of limiting magnitudes or whether one would benefit from combining them. The two methods 
may differ not only in terms of their performance in finding candidates, but also in terms of the characteristics
of the resulting lensed supernova samples. This in turn raises the question which supernova sample and thus 
which method would be suitable for robust measurements of time delays for the purpose of cosmological 
inference. In order to address these issues, we compute detection rates for both methods and a 
wide range of possible filters and supernova types. This allows us to make the first comprehensive comparison 
of detection methods in terms of the discovery potential and the cosmological constraining power of the expected 
gravitationally lensed supernova samples. We explore the possibilities of boosting discovery rates by combining 
the methods and we revisit the estimates of detections rates for ongoing and upcoming transient surveys.

Estimating discovery rates of gravitationally lensed supernovae is not merely a means for quantifying the efficiency of detection methods. Lensed supernovae can be also regarded as a cosmological tool probing a wide range of physical properties across cosmic time, e.g.\ volumetric rates and luminosity functions of supernovae, lens models or cosmological parameters. Observed detection rates can potentially shed light on some aspects of our current models.

The paper is organized as follows. Section 2 outlines detection methods and the potential of increasing 
their discovery potential by combining them. We describe the computation of discovery rates as well as the
underlying lens model and volumetric supernova rates in Section 3. The results are presented in Section 4 followed 
by a discussion in Section 5 and concluding remarks in Section 6.We adopt a flat $\Lambda$CDM cosmological model 
with $\Omega_{\rm m}=0.3$ and $H_{0}=73\;{\rm km}\;{\rm s}^{-1}\;{\rm Mpc}^{-1}$. 

\section{Observational strategies}

The two main features through which strong lensing manifests itself are the multiplicity of images and the magnification
of each image. Therefore, strongly lensed supernovae can be detected and distinguished from ordinary non-lensed supernovae 
if they appear as spatially and temporarily coincident transients \citep{Ogu2010a} or exceptionally bright 
transients \citep{Gold2017}. In the following, we outline each observational strategy and the related detection criteria 
in more detail.

\subsection{Image multiplicity}

The critical factors for detecting multiple images are the image separations relative to the seeing, the flux contrast between 
the images 
and the apparent magnitudes of the faintest images in the configuration. As a base model for this strategy we adopt detection 
criteria optimized for the Large Synoptic Sky Survey (LSST), as proposed by \citet{Ogu2010a}. Following this approach, a transient 
is identified as a strongly lensed supernova if \textit{i}. the maximum image separation $\theta_{\rm max}$ between images 
falls into a range between $0.5''$ and $4''$, where these limits stem from seeing conditions as well as the choice of selecting 
systems lensed by isolated galaxies, characterized by relatively simple lens models, 
\textit{ii.} the flux ratio between 
the images for doubly imaged supernovae is larger than 0.1 and \textit{iii.} at least three or two images are detected 
for quads/cusps (four/three images) and doubles (two images), respectively. For sufficiently bright 
supernovae and a suitable survey cadence, this strategy not only yields detections of multiply imaged supernovae, 
but it also enables the measurement of basic lensing properties such as time delays and flux ratios between the images. 
In this respect, this is an ideal observational strategy for large and self-contained surveys such as the LSST. 
Initial information on lensing configurations can be also used to pin down the best candidates for follow-up observations.

\subsection{Magnification}

Strongly lensed supernovae can be detected as transients which appear significantly brighter than the brightest supernovae 
at the redshift of the apparent host galaxy (lens galaxy for lensed supernovae or actual host galaxy for non-lensed transients). Following \citet{Gold2017}, the detection criterion can by formulated in the following way:
\begin{eqnarray}
m_{\rm X}(t_{\rm peak}) & < & \langle M_{\rm X}\rangle(t_{\rm peak})+\mu(z_{\rm host})+\nonumber \\
 & & K_{\rm XX}(z_{\rm host},t_{\rm peak})+\Delta m,
\label{criterion_magnification}
\end{eqnarray}
where $m_{\rm X}(t_{\rm peak})$ is the observed peak magnitude of the transient in band $X$, $\langle M_{\rm X}\rangle(t_{\rm peak})$ 
is a mean absolute magnitude of a reference class of brightest supernovae in band $X$ at peak, $\mu$ is the distance modulus, $z_{\rm host}$ 
is redshift of the apparent host galaxy, $K_{XX}(z_{\rm host},t_{\rm peak})$ is a K-correction for the assumed 
reference class of brightest supernovae at the peak of their light curves and $\Delta m<0$ is a free parameter defining the 
magnitude gap between lensed and non-lensed supernovae. Absolute magnitudes and K-corrections can be calculated from spectral templates of the assumed 
reference class of bright supernovae, while the redshift of the apparent host galaxy can be estimated from existing
photometric data from wide-field surveys. Since only rare cases of supernovae are brighter than SNe Ia, 
e.g.\ superluminous supernovae, it is reasonable to use type Ia as the reference class of bright supernovae. Like
\citet{Gold2017}, we assume a mean peak absolute magnitude of $-19.3$ in the $B$ band.

The choice of the parameter $\Delta m$ is dictated by a trade-off between the completeness and the contamination of the selected 
lensed supernova candidates. As a base model, we use $\Delta m=-0.7$ required for distinguishing between the brightest non-lensed type Ia supernovae and gravitationally 
lensed supernovae \citep{Gold2017}. We expect that the assumed magnitude gap is a sufficiently conservative choice from the 
point of view of minimizing the false positive rate. However, one has to bear in mind that rare luminous supernovae brighter 
than $-20$ in the $B$ band inevitably will be confused with lensed supernovae in this approach.

The effective depth of a transient survey in the context of detecting strongly lensed supernovae via magnification is 
modulated by the extent to which all lensing images contribute to the measured flux. It is natural to consider here two scenarios in which the total observed flux either comes from all images or is simply approximated by the flux of the brightest image. The former, which we adopt as our base model, sets an upper limit on the effective 
depth and discovery rates with respect to how efficiently the measured flux is integrated over all images. The latter determines the corresponding lower limits. We expect that most realistic detections in typical transient surveys will be intermediate between the two extremes.

Compared to the strategy based on detection of multiple images, the competitiveness of this method 
heavily relies on follow-up observations. Higher resolution and deeper observations are necessary to both confirm the lensing 
nature of the candidates (by means of detecting multiple images) and determining the lensing configuration.

Although initially envisaged for finding strongly lensed type Ia supernovae, the method can be also applied to detecting gravitationally 
lensed core-collapse supernovae \citep{Gold2018}. Here, using type Ia supernovae as the reference class of bright supernovae 
sets a relatively conservative detection threshold. However, as we shall see, the higher volumetric rates of 
core-collapse supernovae can compensate their lower luminosities leading to even higher discovery rates than for type Ia.

\subsection{Hybrid approach}
As we shall see in Section 4, the two methods described perform very differently. The methods are complementary 
in many respects in that they maximize their efficiencies in 
different ranges of survey depth. This provides motivation for considering a third approach which combines 
the ideas underlying the two techniques. Here, a transient is classified as a lensed supernova candidate if at least 
one of the two methods identifies it as a potential lensed supernova.

\section{Simulations}
We employ a Monte Carlo approach to compute the expected number of observed strongly lensed supernovae. The method relies 
on drawing random realizations of lens galaxies and supernovae in a light cone, and counting strong lensing events which satisfy 
certain detection conditions.

\subsection{Lens galaxies}

Motivated by its success in modeling multiply imaged QSOs, we assume that the mass distribution in the lens galaxies  is adequately represented by a 
Singular Isothermal Ellipsoid (SIE) model \citep{Kor1994} in which the convergence $\kappa$ is given by:
\begin{eqnarray}
\kappa(x,y) & = & \frac{\theta_{E}}{2}\frac{\lambda(e)}{\sqrt{(1-e)^{-1}x^2+(1-e)y^2}},\\
\theta_{E} & = & 4\pi \Big(\frac{\sigma}{c}\Big)^{2}\frac{D_{\rm ls}}{D_{\rm s}},
\end{eqnarray}
where $\theta_{E}$ is the Einstein radius, $\sigma$ is the line-of-sight velocity dispersion of the lens galaxy, $D_{\rm s}$ and 
$D_{\rm ls}$ are angular diameter distances, respectively, between the observer and the source, and the lens galaxy and the source. 
The convergence depends on the shape of the lens galaxy through the ellipticity $e$ of the projected lensing mass surface density, 
which includes the contributions from dark matter and baryonic components, both relevant for the statistics of strong 
lensing images \citep{Cas2018}. 
We assume that the ellipticity is given by the light distribution. Based on observational constraints, we adopt a Gaussian distribution 
for the ellipticity with a mean of 0.3 and dispersion of 0.16, with a truncation at 0.1 and 0.9 \citep{Ogu2008}. The function $\lambda(e)$ 
is the so-called dynamical normalization and it depends on the deprojected shape of the lens galaxy. Following \citet{Chae2003}, 
we assume that both oblate and prolate ellipsoids approximate the actual shapes of lens galaxies with equal probabilities, implying 
$\lambda(e)\approx1$.

In order to make the simulated lensing more realistic, we account for the effect of the lens environment by including external shear \citep{Koch1991,Witt1997,Kee1997} with a potential given by 
\begin{equation}
V(x,y)=\frac{\gamma}{2}(x^{2}-y^{2})\cos(2\theta_{\gamma})+\gamma xy\sin(2\theta_{\gamma}),
\end{equation}
where $\gamma$ is the magnitude of the external shear and $\theta_{\gamma}$ is its position angle on the image plane. 
We assume that $\gamma$ follows a log-normal distribution with mean 0.05 and dispersion 0.2 dex, as expected for the external shear around early-type galaxies, calculated by ray tracing in cosmological simulations of the standard cosmological model  \citep{Hol2003}. We expect that the employed simulation-based calibration of the external shear to a large extent accounts for lensing effects from realistic structures around lens elliptical galaxies, although the presence of a galaxy cluster may in addition affect the image configuration. The external shear in our model is uncorrelated with the orientation of the lens galaxy, i.e.\ it has a random orientation in the image plane.

We model the mass function of lens galaxies in terms of the velocity dispersion function of early-type galaxies, which are most common 
lens galaxies. The velocity dispersion function is well approximated by a modified Schechter function of the following form:
\begin{equation}
\frac{dn}{d\sigma}=\phi_{\star}\Big(\frac{\sigma}{\sigma_{\star}}\Big)^{\alpha}
\exp\Big[-\Big(\frac{\sigma}{\sigma_{\star}}\Big)^{\beta}\Big]\frac{\beta}{\Gamma(\alpha/\beta)}\frac{1}{\sigma},
\end{equation}
where $n$ is the comoving density of galaxies. We use the above probability density to draw random realizations of lens galaxies in our 
calculations. We adopt parameters derived from fitting this model to the SDSS data \citep{Choi2007}, i.e.\
\begin{eqnarray}
\phi_{\star} & = & 8.0\times10^{-3}h^{3}\rm{Mpc}^{-3} \nonumber \\ 
\sigma_{\star} & = & 161\,\rm{km}\,\rm{s}^{-1} \nonumber \\
\alpha & = & 2.32 \nonumber \\
\beta & = & 2.67.
\end{eqnarray}
Following \citet{Gold2017}, we narrow the range of velocity dispersions to $(50\,\rm{km}\,\rm{s}^{-1}, 400\,\rm{km}\,\rm{s}^{-1})$ 
for galaxies which can effectively act as gravitational lenses. The lower limit coincides also with the minimum velocity 
dispersion measured from the SDSS spectroscopic observations.

We assume that the velocity dispersion function does not evolve with redshift. Although observational constraints on the velocity dispersion function are limited at high redshifts, this assumption is corroborated by existing data showing no evidence of redshift evolution at $z\lesssim1$, especially for the high velocity dispersion tail \citep{Mont2017,Bez2011}. The assumption is also supported by theoretical arguments based on the standard model for halo formation, for which \citet{Mitch2005} showed that the normalization of the velocity dispersion function can be higher only by $\sim15$ per cent at redshift $z=1$, at which only rare galaxies can generate strongly lensed images of high-redshift supernovae detectable in surveys with depths comparable to the LSST.

The number of lens galaxies in an observational cone as a function of redshift and velocity dispersion is given by
\begin{equation}
\frac{dN_{\rm lens}}{d\sigma dz}=4\pi \frac{c}{H_{0}}\frac{(1+z)^{2}D_{\rm A}^{2}}{E(z)}\phi(\sigma),
\label{number_lens}
\end{equation}
where $D_{\rm A}$ is the angular diameter distance, and $E(z)=H(z)/H_{\rm 0}$ is the dimensionless Hubble parameter. 
We note that the simulated redshift distribution of lens galaxies is independent of the Hubble constant adopted in our work 
and the only role of the assumed cosmological model is to extrapolate the redshift distribution from the 
redshifts of the main galaxy sample in the SDSS to higher redshifts.

\subsection{Supernovae}
The total number of observed supernovae per unit time as a function of redshift is given by
\begin{equation}
\frac{dN_{\rm SN}}{dz}=4\pi \frac{c}{H_{0}}\frac{(1+z)^{2}D_{\rm A}^{2}}{E(z)}\frac{1}{1+z}n_{\rm SN}(z),
\end{equation}
where $n_{\rm SN}(z)$ is the volumetric supernova rate in a local rest frame. Compared to eq.~(\ref{number_lens}), 
the additional factor $1/(1+z)$ accounts for a conversion from a local to the observer rest frame.

\begin{figure}
\centering
\includegraphics[width=0.48\textwidth]{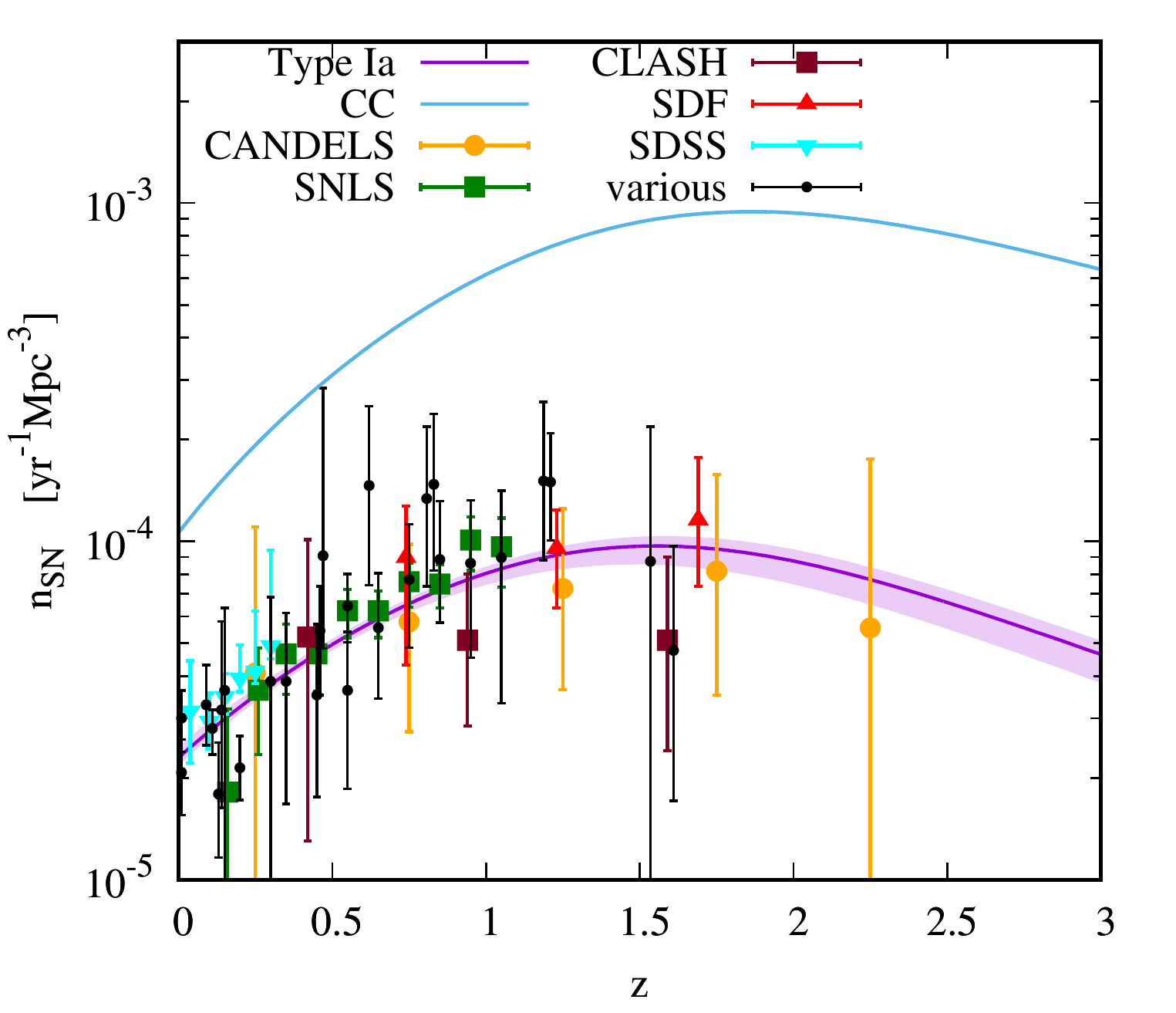}
\caption{Comoving volumetric rate of type Ia and core-collapse supernovae assumed in our study. For core-collapse type, 
the rate is given by the star formation history from \citet{Mad2014}; for type Ia, the rate is the convolution of a delay 
time distribution given by eq.~(\ref{dtd}) with the adopted model of star formation history. Data points show observational 
measurements of the rate for type Ia, as compiled by \citet{Gra2014}. The various distinct symbols highlight selected results from 
CANDELS \citep{Rod2014}, 
the Supernova Legacy Survey \citep[SNLS][]{Per2012}, the Cluster Lensing And Supernova survey with Hubble \citep[CLASH][]{Gra2014}, 
the Subaru Deep Field \citep[SDF][]{Gra2011} and the Sloan Digital Sky Survey \citep[SDSS][]{Dil2010}. 
Black symbols include results from 
\citet{Gra2013,Bar2012,Mel2012,Li2011,Rod2010,Bot2008,Dah2008,Dil2008,Hor2008,Bla2004,Ton2003,Pai2002,Har2000,Cap1999}.
}
\label{sn_rate}
\end{figure}

We compute the volumetric type Ia supernova rate as the convolution of the delay time distribution $\rm{DTD}(t)$ with the 
star formation history $\psi(t)$, i.e.
\begin{equation}
n_{\rm SNIa}(t)=\int_{0}^{t}\psi(t-\tau)\rm{DTD}(\tau)\textrm{d}\tau.
\end{equation}
For the star formation history, we employ a model from \citet{Mad2014}, based on a compilation of the cosmic star formation rates 
determined from UV and IR observations:
\begin{equation}
\psi(z)=0.015\frac{(1+z)^{2.7}}{1+[(1+z)/2.9]^{5.6}}\;M_{\odot}\;\rm{yr}^{-1}\;h_{\rm 70}\;\rm{Mpc}^{-3}.
\end{equation}
Following \citet{Rod2014}, we assume that the delay time distribution is accurately described by the following piecewise function:
\begin{equation}
\rm{DTD}(t)= \left\{ \begin{array}{ll}
0 & \rm{if}\;\;\;t<0.04\;\rm{Gyr} \\
\eta_{1} & \rm{if}\;\;\;0.5\;\rm{Gyr}<t<0.04\;\rm{Gyr} \\
\eta_{0}t^{-1} & \rm{if}\;\;\;t>0.5\;\rm{Gyr}
\end{array}\right.
\label{dtd}
\end{equation}
where $\eta_{0}$ and $\eta_{1}$ are free parameters. This model is well motivated by observations \citep{Rod2014,And2018} 
as well as theoretical arguments derived from the evolution of binary systems 
\citep{Mao2012}. It accounts both for the population of prompt supernovae with delay time 
$t<0.5\;\rm{Gyr}$ \citep[with $40\;\rm{Myr}$ as the shortest possible time before explosion,][]{Bel2005}
and delayed supernovae with $\rm{DTD}\propto t^{-1}$ and $t>0.5\;\rm{Gyr}$. The two free parameters can be determined by fitting the model to the rates inferred from observations. 
We carry out the fit using a compilation of observational determinations of 
the volumetric type Ia supernova rates from \citep{Gra2014} updated with the results from the Cosmic Assembly 
Near-infrared Deep Extragalactic Legacy Survey \citep[CANDLES;][]{Rod2014}. Minimization of $\chi^{2}$ with errors 
including statistical and systematic uncertainties yields 
$\eta_{0}=1.02_{-0.15}^{+0.27}\times10^{-4}\;h_{70}^{2}\;\rm{yr}^{-1}\;\Msun^{-1}$ 
and $\eta_{1}/\eta_{0}=11.98^{+3.37}_{-4.49}$. Fig.~\ref{sn_rate} compares the resulting best fit model 
to the observational data. The obtained constraints imply a fraction of prompt supernovae 
$f_{\rm p}=0.63_{-0.11}^{+0.07}$, consistent with the results obtained by \citet{Rod2014}. 

The rate of core-collapse supernovae is directly proportional to the star formation history $\psi(z)$:
\begin{equation}
n_{\rm SNCC}(z)=k_{\rm CC}\psi(z),
\end{equation}
where $k_{\rm CC}$ is the number of stars that explode as supernovae per unit mass. For our study, 
we adopt $k_{\rm CC}=0.0068\Msun^{-1}$ expected for a mass range of supernova progenitors 
$(8\;M_{\odot},\;40\;M_{\odot})$ and a Salpeter initial mass function. Since the same initial function 
was consistently assumed in the derivation of the star formation rate from observations, the predicted 
rate of core-collapse supernovae is practically independent of the initial mass function \citep{Mad2014}. 
The resulting core-collapse supernova rate is shown in Fig.~\ref{sn_rate}.

\begin{figure}
\centering
\includegraphics[width=0.48\textwidth]{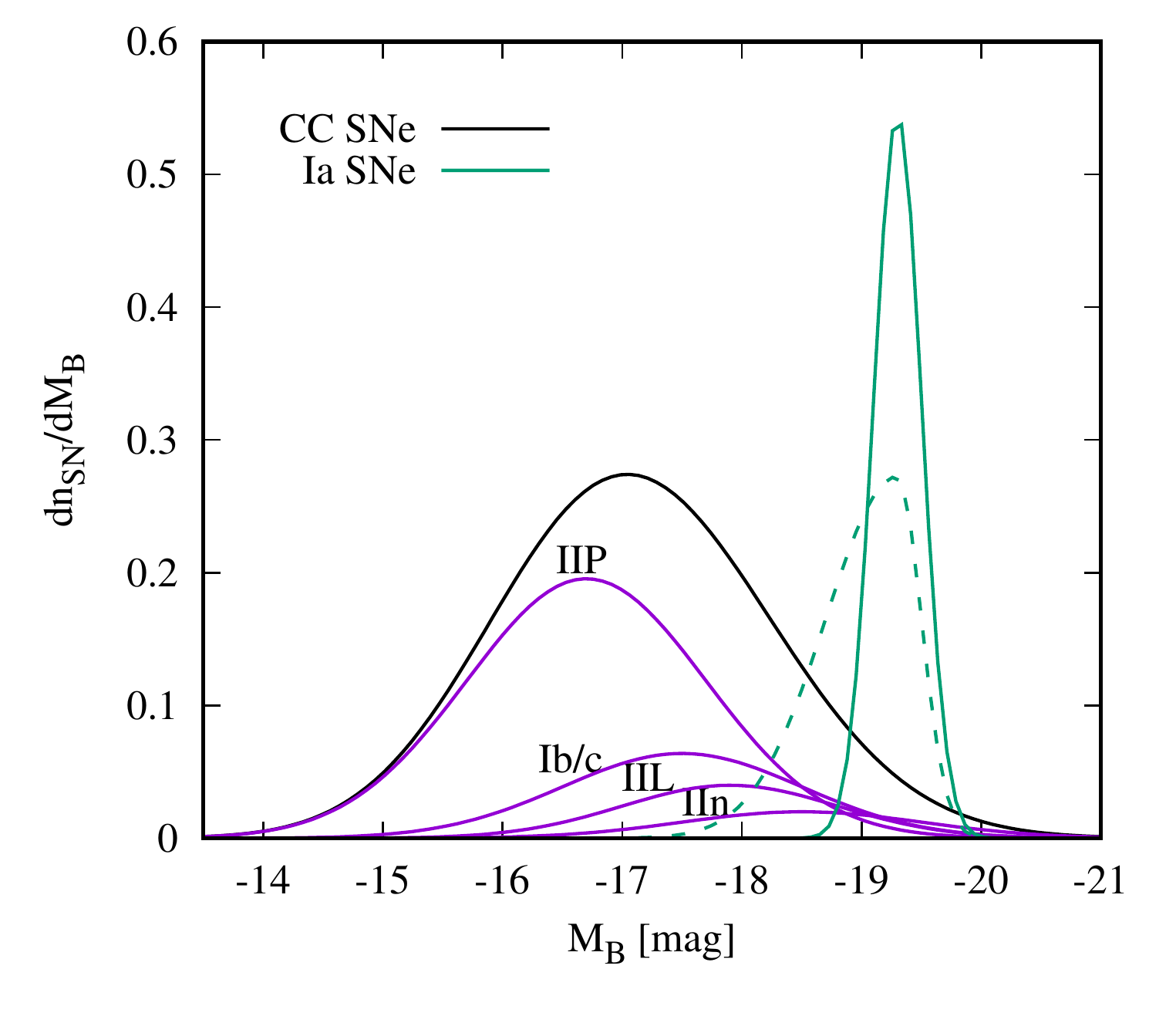}
\caption{Distribution of absolute luminosities in the $B$ band of the different types of supernovae 
considered in our study. Normalizations of the distributions are proportional to the local volumetric rates of the correposnding 
supernova classes. The green dashed curve shows an alternative  
model for type Ia supernovae with with a tail accounting for low-luminosity supernovae.}
\label{abs_lum}
\end{figure}

We calculate light curves and K-corrections of simulated type Ia supernovae (normal branch) using a time series of spectral templates 
computed by \citet{Nug2002}. Following \citet{Gold2017}, we assume that the absolute magnitude is normally distributed 
with a mean of $-19.3$ in the $B$ band and a scatter of $0.2$. We also check the impact of using a more realistic 
distribution with 3 times longer Gaussian tail at low luminosities accounting for faint supernovae 
in volume-limited samples \citep[see e.g.][]{Li2011,Gold2017}.
For core collapse supernovae, we consider types Ib/c, IIP, IIL 
and IIn, assuming that their relative contributions to the total number 
density of core collapse supernovae are independent of redshift and 
constrained by low-redshift observations resulting in 16, 49, 9 and 5 per cent 
respectively for Ib/c, IIP, IIL and IIn \citep{Gra2017,Li2011}. 
We realize light curves of the four supernova subclasses using spectral templates from a compilation 
which is an extension of the work by \citet{Nug2002}
\footnote{https://c3.lbl.gov/nugent/nugent\_templates.html}, based primarily on data from \citet{Lev2005} for type Ibc,  
\citet{DiC2002} for IIn and \citet{Gil1999} for the remaining two types. We approximate the distribution of absolute magnitudes in the
$B$ band by Gaussians with 
mean and scatter of $-17.50$ and 1.0 for type Ib/c, $-16.70$ and 1.0 for type IIP, $-17.9$ and 0.90 for type IIL, 
$-18.5$ and 1.0 (after discarding two outliers with $M_{B}=-15.1,-22.2$) for IIn \citep[][for $h=0.73$]{Rich2014}. The assumed luminosity functions for all classes of supernovae are shown in Fig.~\ref{abs_lum}. 
We note that the borderline between different classes may
not be entirely clear cut. For example, there may be a continuous
transition between type IIP and type IIL supernovae \citep[e.g.,][]{2014ApJ...786...67A}. In this sense, these types can be seen as being representative 
for the fainter and more luminous ends of the type II supernova population.

\subsection{Computation}

First, we generate a random sample of lens galaxies at redshifts $z<1.2$. We find that it is sufficient to realize 
$10^{5}$ lens galaxies and then rescale the final counts of strongly lensed supernovae according to the actual number of 
lens galaxies contained in the assumed comoving volume. Then, we populate the observational cone up to redshift $z=3$ with 
supernovae. In order to reduce shot noise, we artificially increase the supernovae rate by a factor of $5\times10^{5}$. The actual 
detection rate is then retrieved by scaling down the counts by the same factor. In order to save computational time (driven 
primarily by lensing computations), we discard all supernovae at angular distances larger than $4\theta_{E}$. These supernovae 
are typically too far from the lenses to be multiply imaged or strongly magnified. Our choice of a maximum redshift for the lenses and 
the supernovae allows us to determine detection rates of strongly lensed supernovae for limiting magnitudes up to 26, which is 
around 2 mag deeper than the envisaged depth for the LSST.

For every lens-supernova pair we calculate all basic lensing properties, i.e.\ image multiplicity, magnifications, positions of the images and 
time delays. The lensing equations are solved numerically using the publicly available code for lensing calculations, \textit{glafic} \citep{Ogu2010}. 
When calculating time dependent apparent magnitudes, we take into account both the effects of magnification and time delay determined 
for every image.

Once the complete Monte Carlo sample of strongly lensed supernovae is computed, we can employ different detection criteria and thus 
determine the number of detectable lensed supernovae as a function of survey depth in several different bands. This part of the calculations is independent of the 
assumed cosmological model and volumetric supernova rates; therefore, it can be repeated for different detection criteria using the same precomputed Monte Carlo sample of strongly lensed supernovae and the corresponding lensing properties. The supernova yields are computed for observations in five SDSS/LSST filters: {\it g, r, i, z} and {\it y}. 

\section{Results}

\subsection{Type Ia supernovae}

\begin{figure}
\centering
\includegraphics[width=0.48\textwidth]{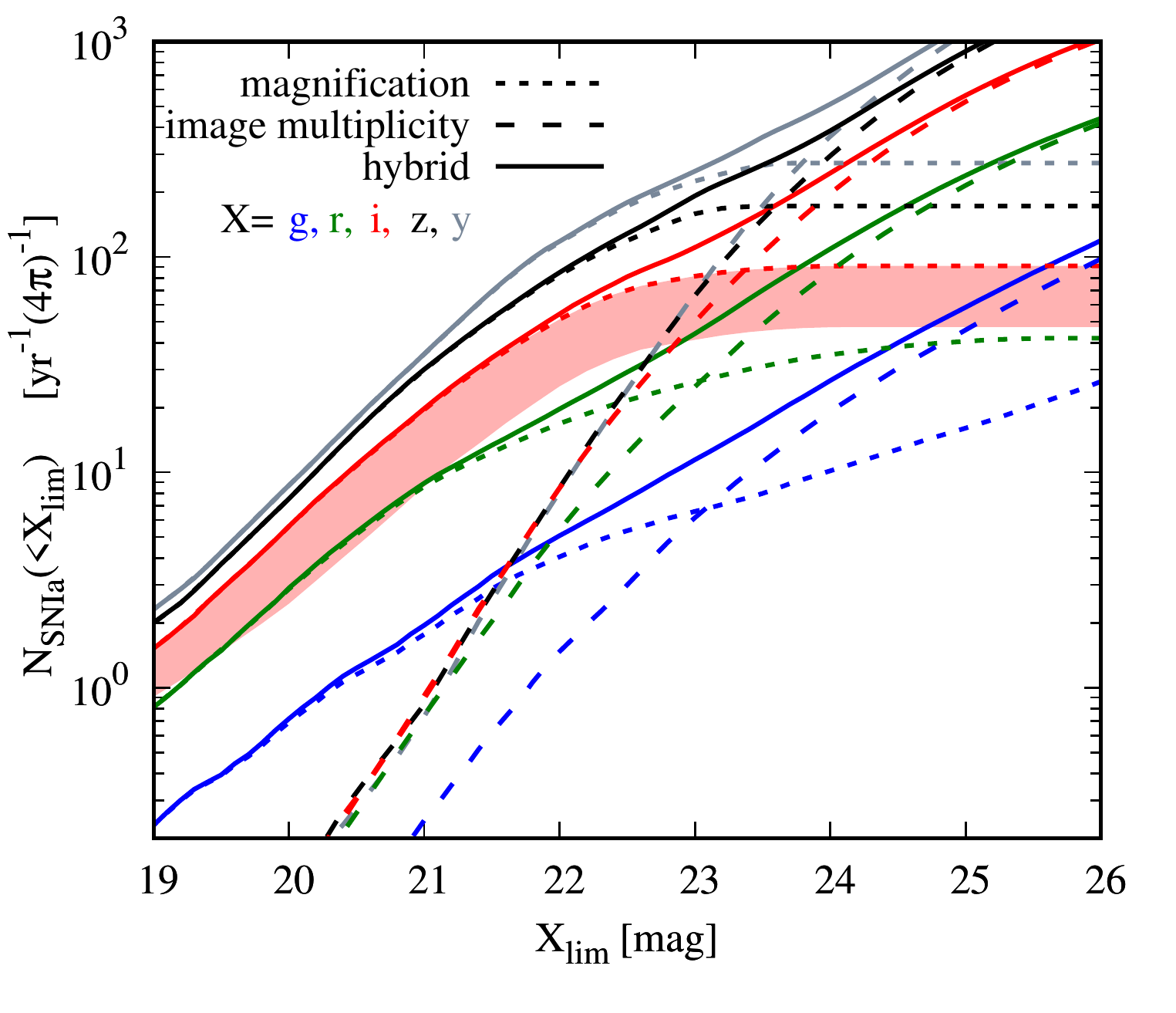}
\caption{Detection rates of strongly lensed type Ia supernovae as a function 
of survey depth in five different bands, in an all-sky search. 
The long and short dashed curves compare the supernovae yields expected for two different observational 
strategies based on detecting image multiplicity or highly magnified supernovae. 
The solid curves show the yields for the hybrid method which maximizes detection rates by means of combining both 
detection criteria. The shaded band indicates the range of detection rates given by the extent to which
flux is integrated over the supernova images, with the upper and lower limits corresponding 
to the total measured flux from all images (upper limit) or solely from the brightest image (lower limit).
}
\label{yields_comparison}
\end{figure}

Fig.~\ref{yields_comparison} compares detection rates for strongly lensed type Ia supernovae expected for the three methods with the base 
choice of free parameters, as outlined in Section 2. Our results demonstrate that the method based on detecting highly magnified supernovae surpasses 
the image multiplicity technique for relatively shallow surveys with limiting magnitudes $\lesssim 22$. In particular, for a survey with a 21 magnitude depth, the magnification method is expected to find $\sim 20$ times more strongly lensed supernovae than the image multiplicity method. Due to the stronger dependence on the limiting magnitude for the image multiplicity method, this ratio becomes 
larger for even shallower surveys, with a 3-fold change per magnitude in the $i$ band.

For limiting magnitudes $\gtrsim 23$, increasing survey depth does not improve the performance of the magnification method, but it appreciably increases 
the rates for the other method. For deep surveys like the LSST, detecting multiple images appears to become a more powerful means 
for finding lensed type Ia supernovae than gravitational magnification. The image multiplicity method is expected to yield about twice
as many detections as the magnification method at a limiting magnitude of $24$. The difference between the two methods becomes even more prominent for deeper surveys: the rates expected for the image multiplicity method increase exponentially with limiting magnitudes in a range between 24 and 26, whereas the rates from the magnification method become constant.

Both methods are expected to find comparable numbers of lensed type Ia supernovae at limiting magnitudes between $23.0$ in the $g$ band to 
$23.8$ in the $y$ band.
The lensed supernova samples returned by the two methods happen to be only weakly overlapping; therefore, combining both detection 
criteria is expected to increase the yields, especially at intermediate limiting magnitudes. 
This is demonstrated by the solid curves which show detection rates for the hybrid method which 
identifies strongly lensed supernovae either as highly magnified or multiply imaged transients. The hybrid method maximizes supernova 
yields at all limiting magnitudes. Unsurprisingly, it follows closely the rates from the magnification method at $X_{\rm lim}\lesssim21$ 
and the image multiplicity technique at $X_{\rm lim}\gtrsim24$.

\begin{figure}
\centering
\includegraphics[width=0.48\textwidth]{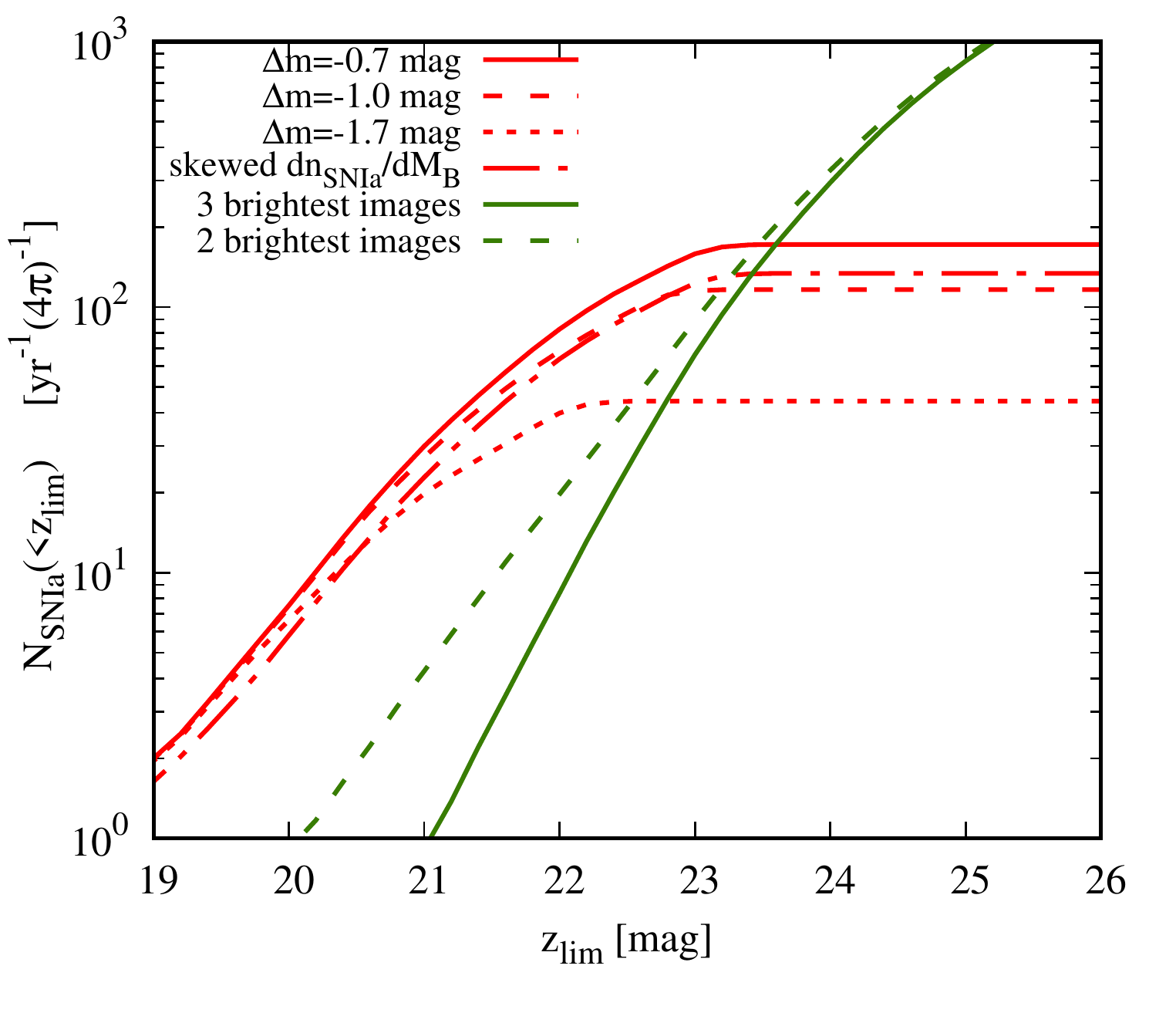}
\caption{Sensitivity of detecting lensed type Ia supernova in the $z$ band to parameters and conditions defining detection criteria 
of the two methods based on selecting highly magnified (red curves) or multiply imaged supernovae (green curves). The parameter 
$\Delta m$ 
is a minimum enhancement of the apparent brightness due to lensing magnification relative to the mean 
magnitude of type Ia supernovae expected in the apparent host (lens) galaxy ($\Delta m=-0.7$ for the base model). 
For the image multiplicity method, discovery rates depend on the minimum number of detectable images. The solid and dashed 
curves compare the two cases for which at least three (base model) or two images are detectable for quads and cusps. 
The red dashed-dotted curves shows the impact of accounting for the possible low-luminosity tail in the luminosity function 
$d\textrm{n}_{\rm SNIa}/dM_{\rm B}$ of type Ia supernovae (see the green dashed curve in Fig.~\ref{abs_lum}) on the 
expected discovery rates. 
}
\label{sn_rate_effects}
\end{figure}

The performance of the magnification method depends on the choice of the number of images that contribute to the total flux in 
the observations. 
The red band in Fig.~\ref{yields_comparison} shows the expected range of the rates modulated by this effect. 
The upper limit corresponds to detections based on flux integrated over all images, whereas the lower limit is expected for detections 
based on flux solely from the brightest image. The average ratio between the upper and lower limits is 2, with a very weak dependence on survey depth and filters. 

The apparent differences between detection rates from the two main methods reflect the fact that gravitational magnification and image multiplicity are not equally 
prominent features of strong lensing at different depths. The relatively poor performances of the magnification method for deep 
surveys ($X_{\rm lim}>23$) or the image multiplicity method for shallower surveys ($X_{\rm lim}<23$) cannot be 
appreciably improved by modifying the criteria for selecting the candidates. We demonstrate these intrinsic limitations in Fig.~\ref{sn_rate_effects}, where we show supernova yields for different choices of the free parameters defining the 
detection criteria in the two methods. Reducing the
magnitude gap $\Delta m$ in the magnification method results naturally in a higher detection rate. 
However, the improvement is limited solely to magnitudes $\gtrsim 22$ where the method is clearly outperformed by the image 
multiplicity technique. Furthermore, this inevitably leads to a higher false positive detection rates due to intrinsically 
bright type Ia supernovae.

\begin{figure}
\centering
\includegraphics[trim={11.5cm 7.0cm 1.cm 7.0cm},clip,width=0.8\textwidth]{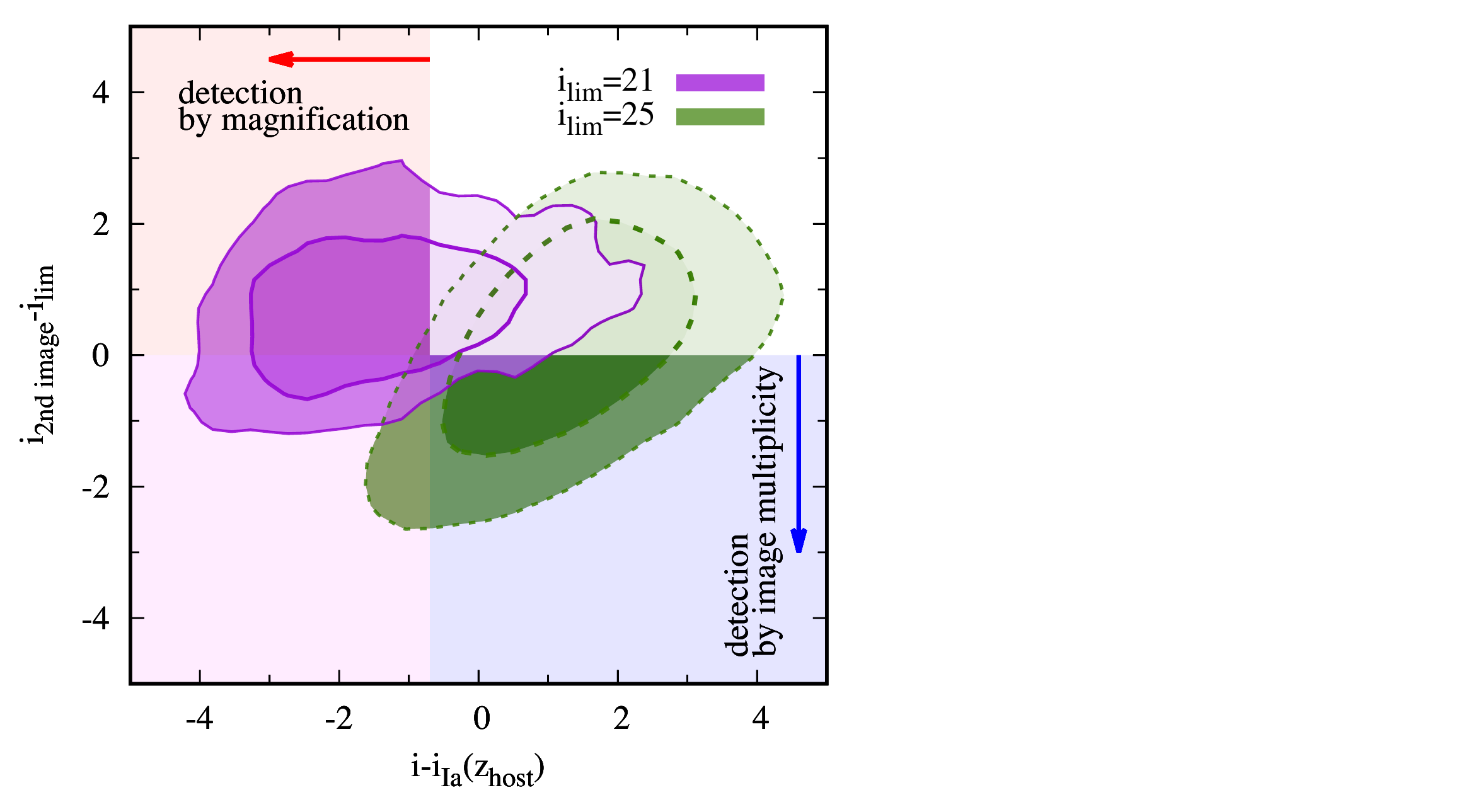}
\caption{Selection of lensed supernova candidates (using type Ia as an example) by means of finding 
highly magnified transients, i.e.\ brighter than a fiducial type Ia supernovae observable in the apparent 
host (actual lens) galaxy (see the red arrow and the indicated shaded region), or detecting the second brightest 
image (see the blue arrow arrow and the indicated shaded region). The two contours show the distribution 
of all observable lensed supernovae in two surveys with depths of 21 and 25 in the $i$ band. The majority of bright, 
low-redshift supernovae appear to be highly magnified, but featuring the secondary image beyond detection limits 
(purple contours). They can be found using primarily the magnification method. On the other hand, the majority of faint, 
high-redshift lensed supernovae from a deep survey (green contours) are not sufficiently magnified to be 
detected as peculiarly bright supernovae, but they exhibit detectable second brightest images. These lensed supernovae 
can be effectively found as doubly imaged transients.
}
\label{selection}
\end{figure}

Relaxing the selection criteria for the image multiplicity method by including all systems with at least two detectable 
images (for all images configurations) improves the supernova yields only at small limiting magnitudes, e.g.\
by a factor of 4 at $z_{\rm lim}\approx 21$ 
(see the green dashed curve in Fig.~\ref{sn_rate_effects}). The resulting detection rates, however, are still smaller than 
those of the magnification method. Since the selection criterion cannot be relaxed even further, this case sets an upper 
limit for the expected detection rates based on image multiplicity. Finally, we find that changing the range of the maximum 
image separation does not appreciably modify the predictions for the lensed supernova yields.

A strong limitation of the magnification method is reflected by a plateau at $X_{\rm lim}\gtrsim 22.5$ which signifies that 
the method does not benefit from increasing survey depth. This feature 
is an unavoidable drawback of the method and it cannot be removed by simply adjusting $\Delta m$ (subject to a 
reasonable condition $\Delta m<0$). It is primarily caused by the fact that high-redshift lenses and supernovae require 
extremely large, and therefore improbable, magnifications in order to fall within a range of detectable fluxes. 
We illustrate this in Fig.~\ref{selection} which compares selections of lensed supernova candidates in shallow ($i_{\rm lim}=21$) and 
deep ($i_{\rm lim}=25$) surveys. Faint, high-redshift supernovae are not sufficiently magnified in order to be selected 
by the magnification method. Most of them happen to be fainter than a fiducial reference type Ia supernova with 
absolute luminosity $M_{B}=-20$ that would be observed in the apparent host galaxy (actual lens galaxy). On the other hand, faint supernovae 
from deep surveys appear to exhibit relatively brighter secondary images than those from shallow surveys. This makes 
the image multiplicity method more effective in selecting lensed supernova candidates in deep surveys at $X_{\rm lim}\gtrsim 23.5$.

\begin{figure}
\centering
\includegraphics[width=0.48\textwidth]{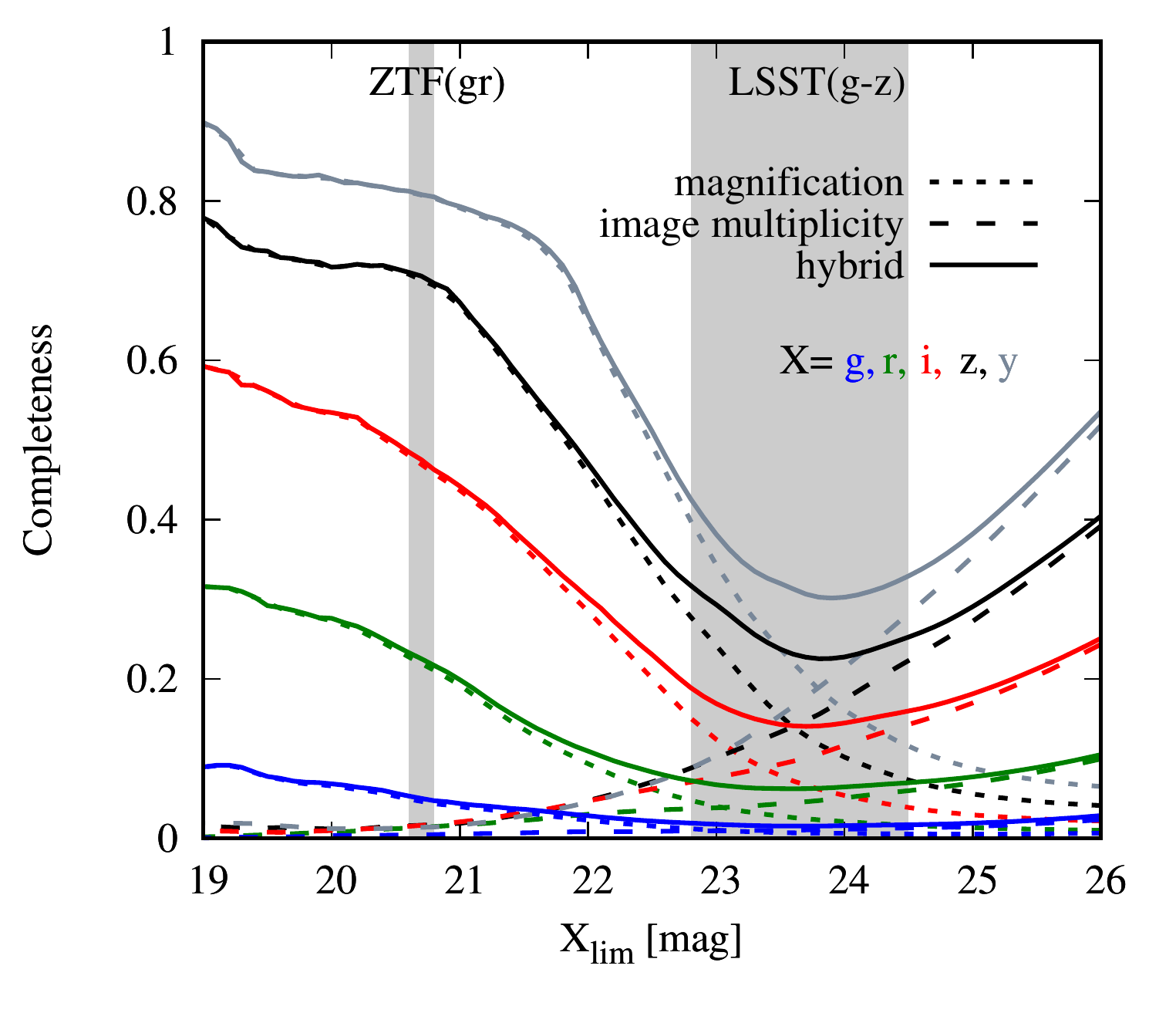}
\caption{Completeness of lensed supernova candidates found with three different methods, as a function 
of survey depth in five different bands. The vertical grey bands indicate the depths of ZTF and LSST. The method based on detecting highly magnified supernovae reaches 
an $\sim80$ per cent completeness in the $z$, $y$ bands at $X_{\rm lim}<21$. The completeness of lensed supernovae 
search by means of detecting image multiplicity increases with survey depth. The method becomes more complete than that based on magnification for surveys with depths comparable to or larger than the LSST. 
}
\label{completeness}
\end{figure}

\begin{figure}
\centering
\includegraphics[width=0.48\textwidth]{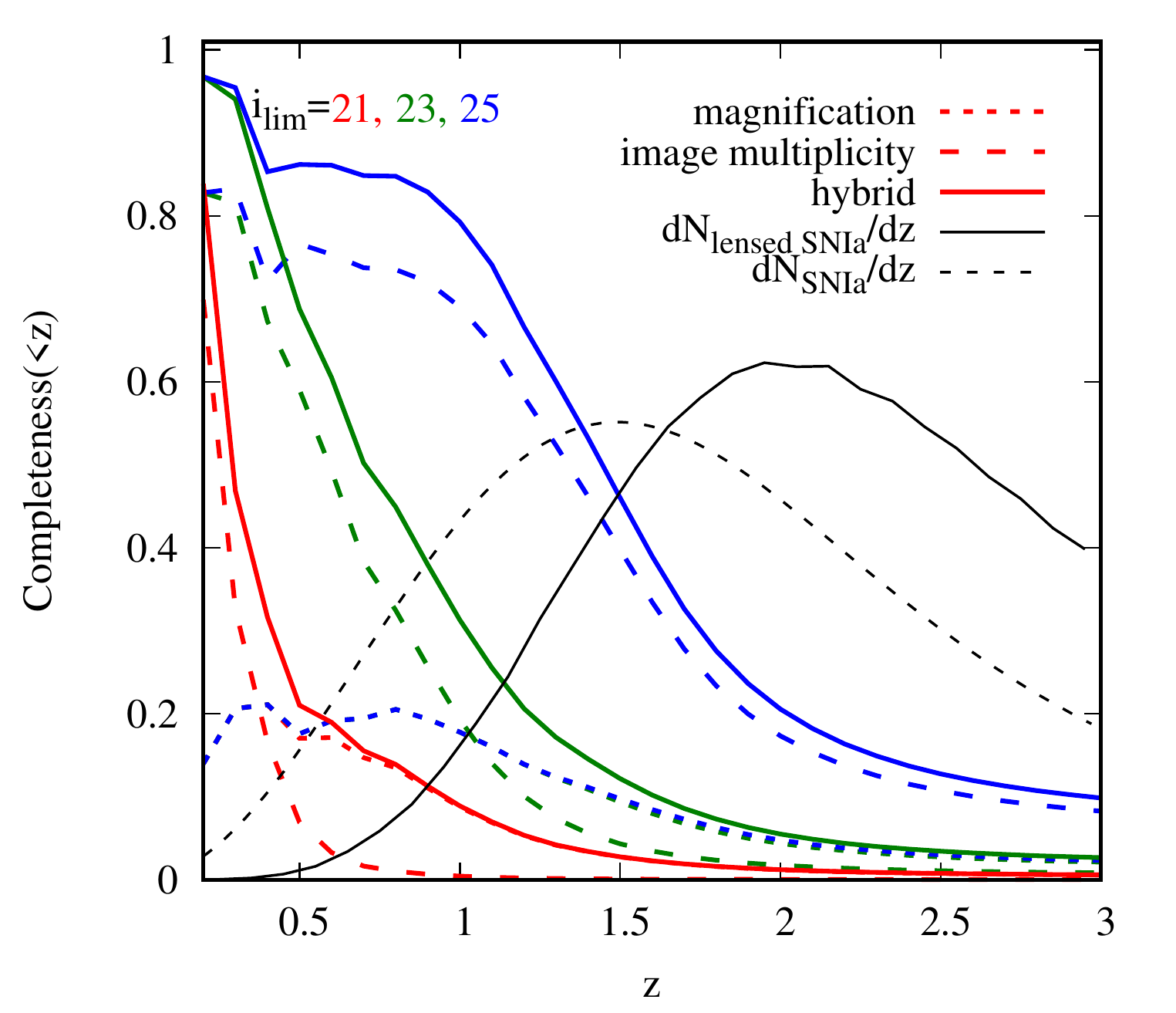}
\caption{Redshift completeness of lensed type Ia supernova candidates found with three different methods, for three limitting magnitudes $i_{\rm lim}$ in the $i$ band. The method based on detecting image multiplicity returns nearly complete redshift samples up to a maximum redshift set by the survey depth. For a reference point, the black curves show redshift distributions of all (non-lensed) or strongly lensed type Ia supernovae, both normalized to 1 at $z<3$.
}
\label{completeness-z}
\end{figure}

\begin{figure*}
\centering
\includegraphics[width=0.48\textwidth]{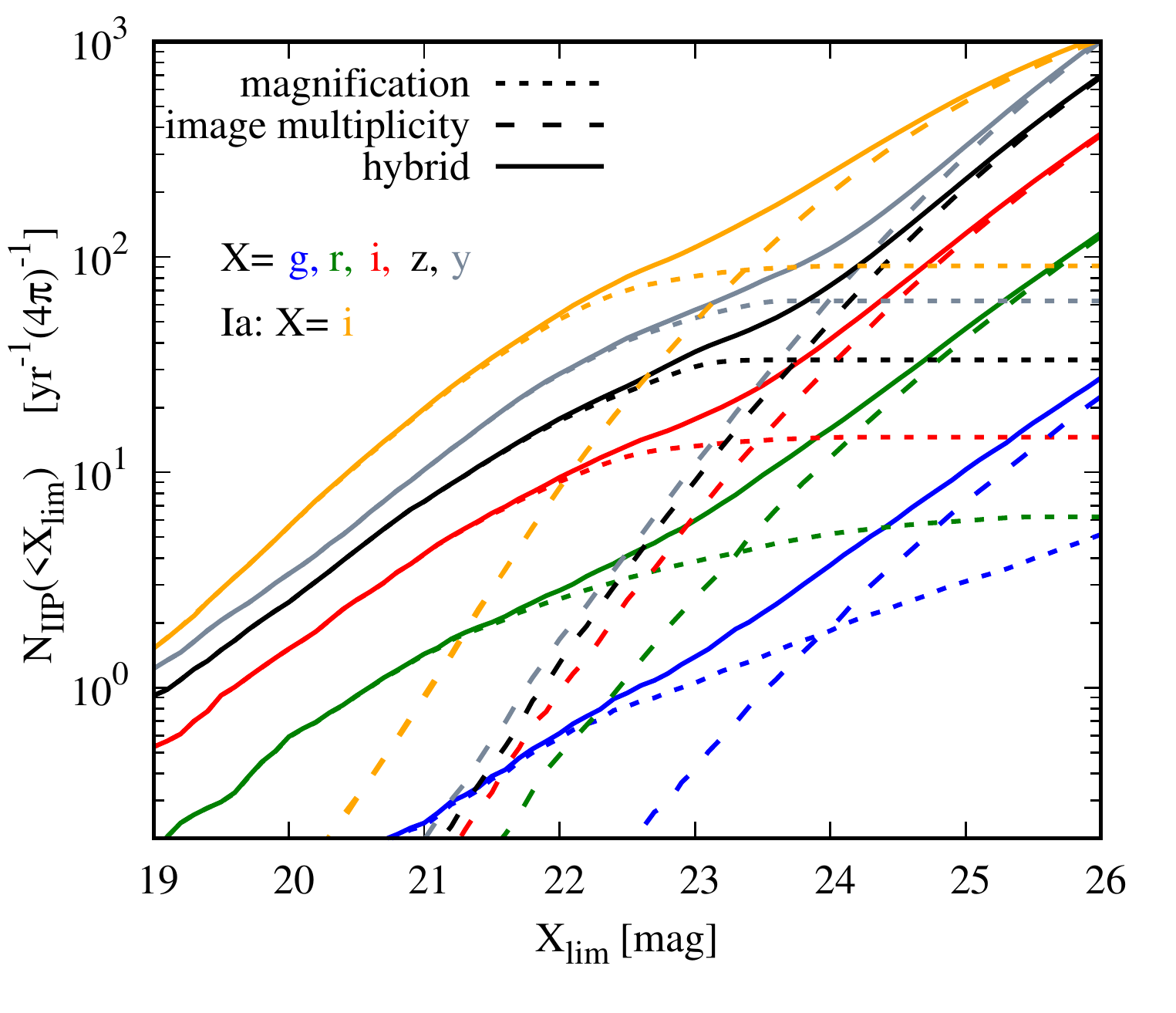}
\includegraphics[width=0.48\textwidth]{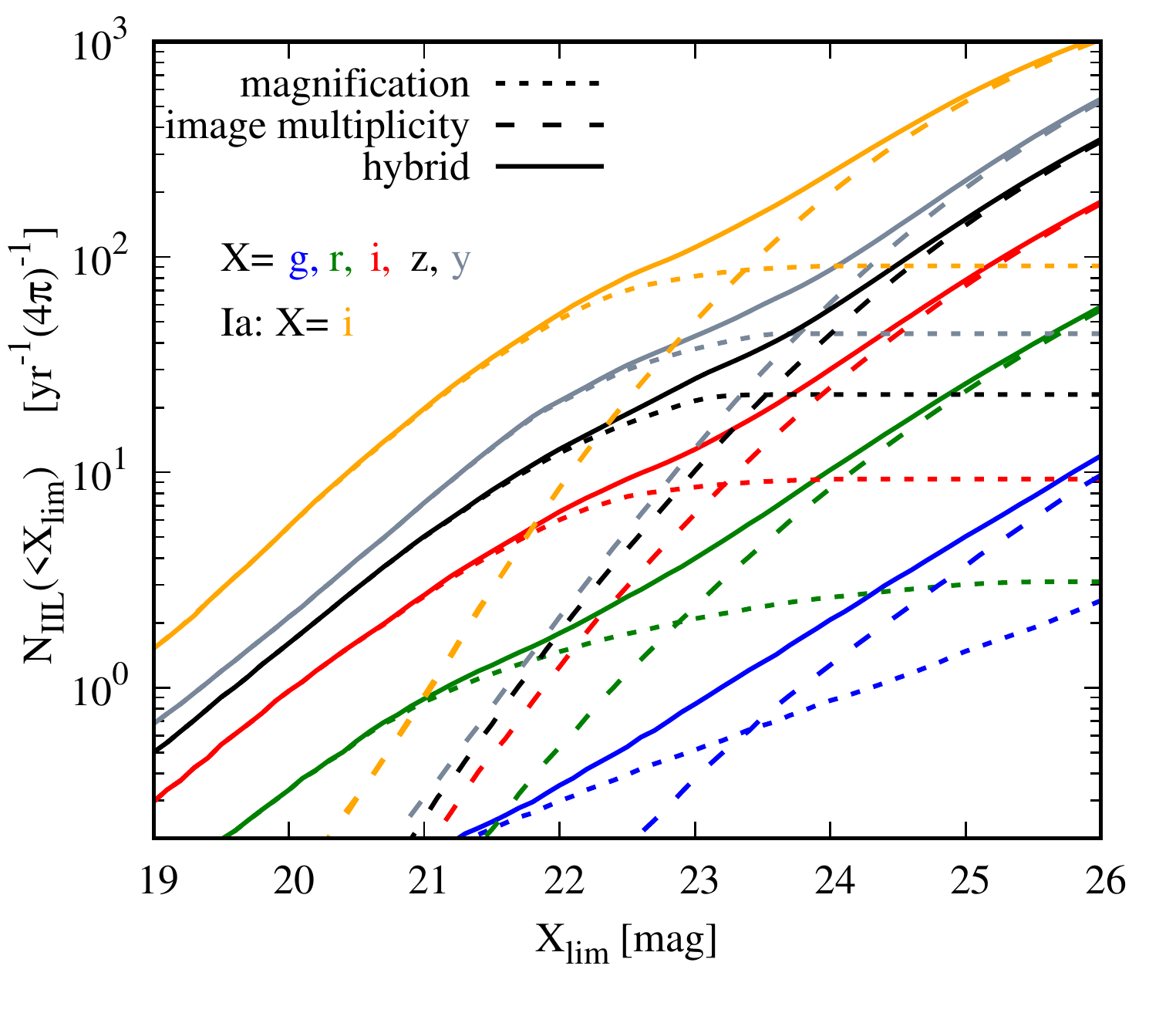}
\includegraphics[width=0.48\textwidth]{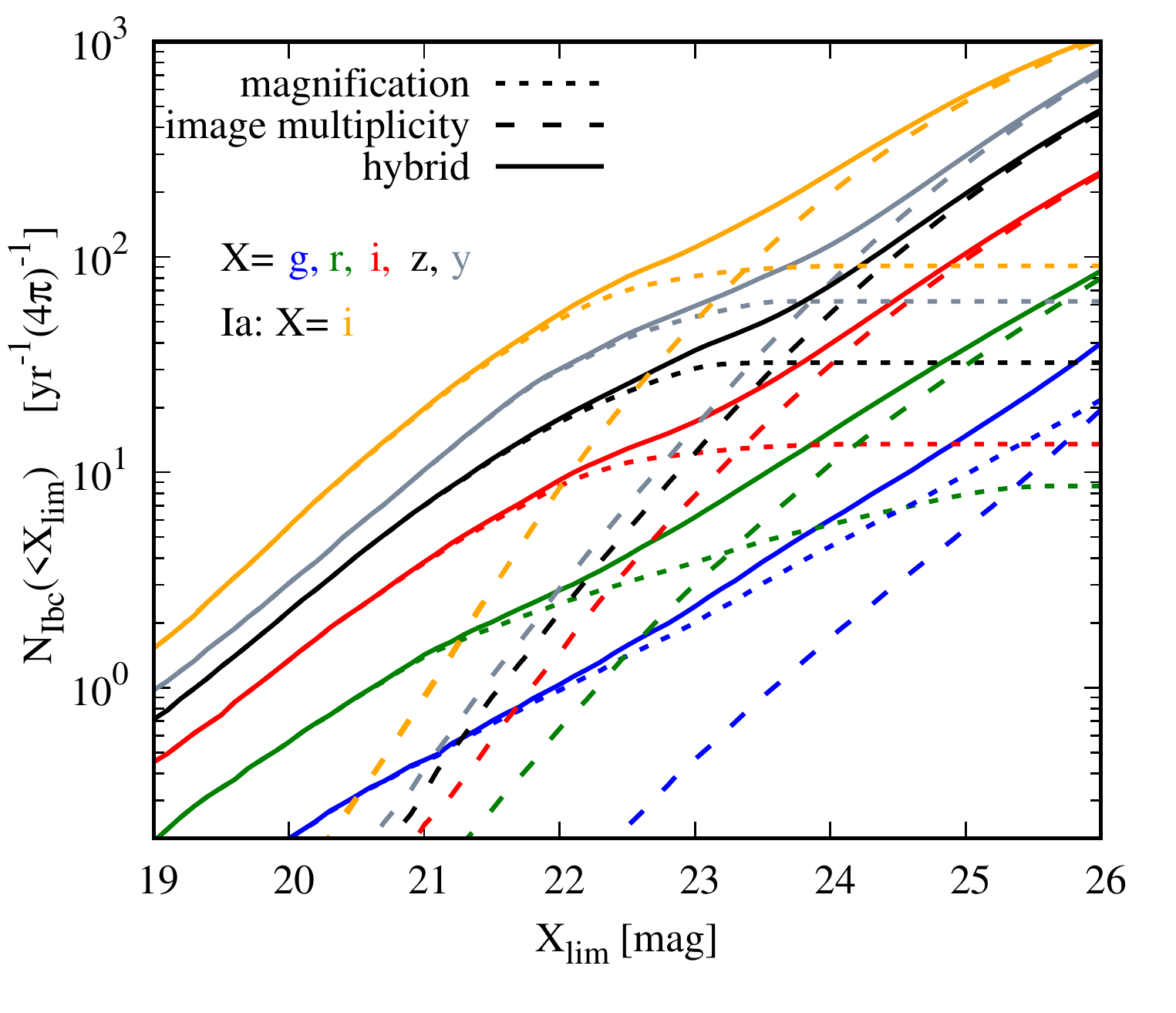}
\includegraphics[width=0.48\textwidth]{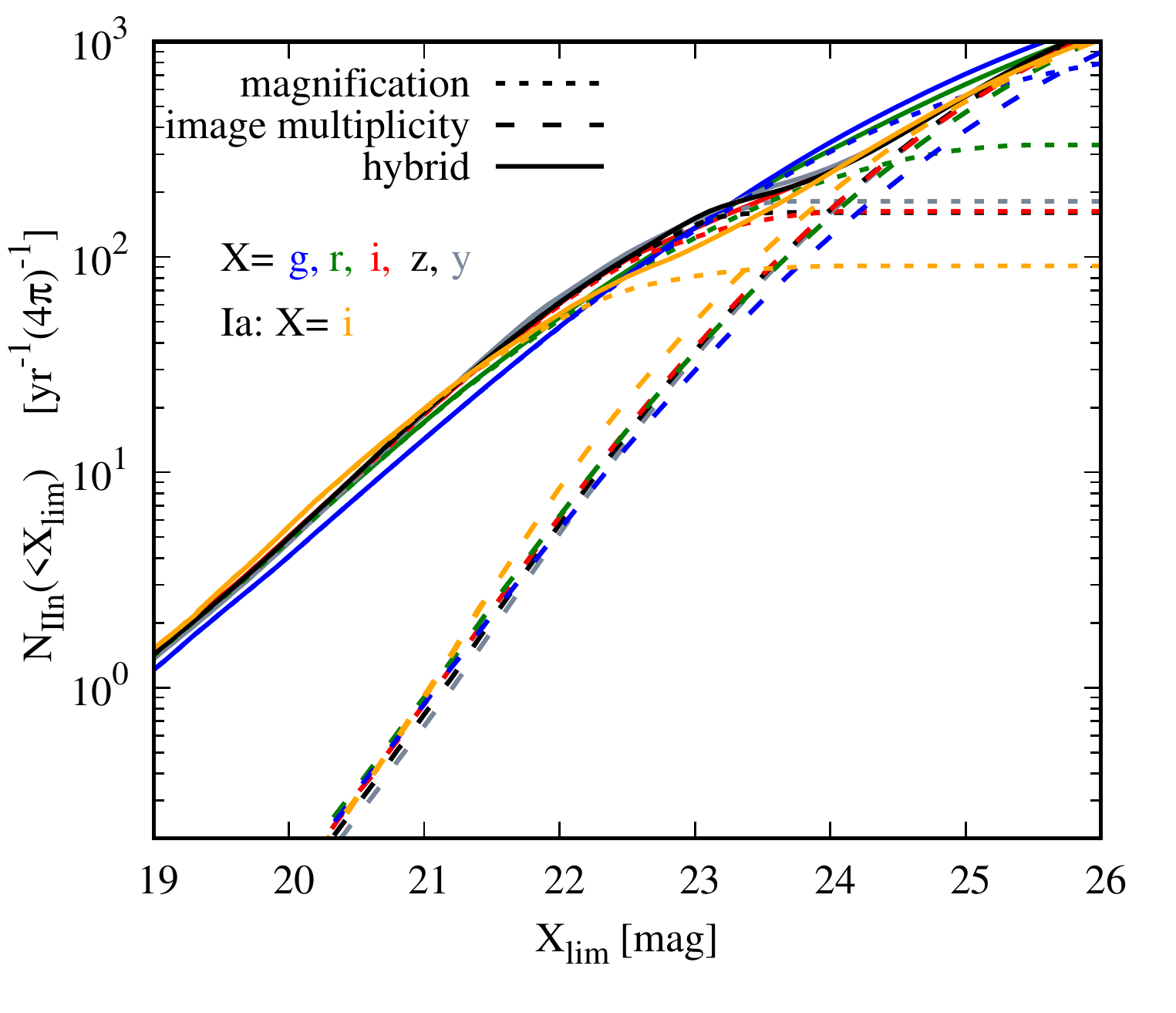}
\caption{Detection rates of strongly lensed core-collapse supernovae (from top left clockwise: 
IIP, IIL, IIn and Ibc) as a function of survey depth in five different bands, in an all-sky search. 
The long and short dashed curves compare the supernova yields expected for two different observational 
strategies based on detecting image multiplicity or highly magnified supernovae. 
The solid curves show the yields for the hybrid method which maximizes detection rates by means of combining both 
detection criteria. For comparison, the orange curves shows the expected rates for lensed type Ia supernovae 
observed in the $i$ band.}
\label{rates-CC}
\end{figure*}

In order to show more quantitatively the differences between the magnification and image multiplicity methods 
of selecting lensed supernovae candidates, we calculate the completeness of lensed supernova searches 
as a function of survey depth (see Fig.~\ref{completeness}). 
We define the 100 percent complete reference sample as consisting 
of all observable/detectable lensed type Ia supernovae, i.e.\ all multiply imaged type Ia 
supernovae that are brighter than $X_{\rm lim}$ at the peak of their light curves in at least one of the five bands, 
where apparent magnitudes are computed by integrating flux over all images. The figure shows that the 
magnification method maximizes its completeness for shallow surveys, with completeness reaching $80$ 
per cent at $\lesssim 21.5$ mag in the $z$ or $y$ band, and then it degrades at large survey depths. An inverse 
trend characterizes the completeness of the image multiplicity method which becomes more complete with increasing 
survey depth, reaching about 40 per cent at 25 mag in the $y$ or $z$ bands. The solid curves show the completeness of the hybrid 
method. It is evident that combining the magnification and the image multiplicity detection criteria improves the completeness 
at limiting magnitudes characteristic for the LSST, i.e.\ $22<X_{\rm lim}<24$. In particular, the hybrid method boosts the completeness 
by about $50$ per cent in the $y$ band relative to lensed supernova searches based solely on the magnification or image multiplicity criterion. 
Figure~\ref{completeness-z} demonstrates also a difference between the magnification and image multiplicity methods in terms 
of redshift completeness. While the magnification technique can never be more complete than 20 per cent at all redshifts, 
the image multiplicity method attains an 80-90 per cent completeness up to a maximum redshift set by the survey depth.

\subsection{Core-collapse supernovae}

In Fig.~\ref{rates-CC}, we show the expected detection rates for lensed core-collapse supernovae, divided into
four subclasses. As for Type Ia supernovae, the method of finding lensed supernovae by means 
of detecting highly magnified transients appears to be more effective for shallow surveys, but it is surpassed by the technique 
based on image multiplicity for deep surveys. The limiting magnitude of comparable performances of the two methods 
depends weakly on the supernovae type and filter, and it typically falls within a magnitude range between 23 and 24. The exception 
in this respect are type IIn supernovae for which the magnification-based method appears 
to be more effective up to as high as 25 magnitude in the $g$ and $r$ bands.

The orange curves in Fig.~\ref{rates-CC} show reference rates for lensed type Ia supernovae detectable 
in the $i$ band. Except for type IIn supernovae, the discovery rates of lensed core-collapsed supernovae are lower than type Ia by a factor 
of 2--5. A high fraction of luminous type IIn supernovae with $M_{\rm B}<-20$ makes them appreciably easier to detect. 
Furthermore, the high UV flux of these supernovae substantially increases the detection rates in shorter wavelength filters. 
This considerably reduces the differences between detection rates in the $g$ and $y$ bands, which is otherwise conspicuous for 
all other supernova types. This effect becomes particularly strong at large limiting magnitudes ($X_{\rm lim}>23.5$), where 
the $g$ band appears to be the most effective filter for detecting lensed type IIn supernovae via the magnification method.

The total number of detectable lensed core-collapse supernovae exceeds that of type Ia, with the primary contribution 
coming solely from type IIn supernovae. The detection rates of lensed supernovae of type IIP, IIL and Ib/c are systematically 
lower than type Ia.

\subsection{Fitting function}

For the purpose of future studies exploring the feasibility of using strongly lensed supernovae as cosmological 
and astrophysical probes, we provide a set of simple fitting functions reproducing the computed detection rates. 
We find that the logarithm of detection rates can be well fitted by a fourth degree polynomial, i.e.
\begin{eqnarray}
\log_{10}n_{\rm SN}(<X_{\rm lim}) & = & \sum_{i=0}^{i=4}X_{i}(X_{\rm lim}-22.5)^{i}.
 \label{detection_model}
\end{eqnarray}
Table~\ref{appro} lists best fit parameters for different detection strategies, supernova types and filters. 

The fitting functions provide accurate approximations to the exact results in a range between $X_{\rm lim}=19$ 
and $X_{\rm lim}=26$ for all cases. The mean precision given by the root mean square averaged over all 
cases of filters, supernova types and methods is 0.02~dex.

\section{Discussion}

The expected detection rates of strongly lensed supernovae computed in our work depend on a range of assumptions. The
cosmological model, the lens model and the volumetric rates of type Ia supernovae are fairly well constrained by observations and 
supported by solid theoretical frameworks. On the other hand, the luminosity functions and fractions of different types of 
supernovae, the volumetric rates of core-collapse supernovae at high redshifts and the spectral templates of different types of core-collapse supernovae (especially type IIn) are less certain and may be a source of additional systematic errors in our 
estimates. Fig.~\ref{sn_rate_effects} shows an example of how modifications in the input luminosity function of type Ia supernovae 
can change the expected discovery rates. Bearing in mind that our results may be subject to improvements in the light of 
future observations, we stress that the predicted discovery rates of lensed supernovae presented here reflect the current state 
of our knowledge on different types of supernovae.

In our calculations we neglect the effect of microlensing. This is a safe assumption, because due to its stochastic nature, microlensing 
is expected to have a negligible impact on overall predictions of discovery rates of lensed supernovae \citep{Gold2018a}. However, 
the effect does perturb light curves independently in every supernova image, giving rise to additional systematic errors in measurements 
of relative fluxes in the images and time delays \citep{Pie2019}. In our comparison of detection methods we do not consider the problem 
of possible false positive detections. All methods will be affected by a population of faint quasars increasing stochastically their 
brightness above the detection limit. Moreover, the magnification method can easily confuse non-lensed superluminous supernovae with 
high-redshift lensed candidates. Quantitative analyses of these effects is worth carrying out in the near future.

In the following, we compare observational strategies for detecting lensed supernovae in the context of ongoing or 
upcoming transient surveys and the potential of using lensed supernovae for cosmological inference.

\subsection{Ongoing and upcoming surveys}

Strongly lensed supernovae will be discovered in appreciable numbers by ongoing and upcoming transient surveys. Table~\ref{rates_survey} 
lists the expected discovery rates for the Zwicky Transient Facility (ZTF), LSST and a hypothetical Pan-STARRS survey.

ZTF is an ongoing survey monitoring 15000~deg$^{2}$ in the $g$ and $r$ bands. For the estimation of discovery rates, we assume that all candidates will be detected in the most effective filter, i.e.\ the $r$ band. A $5\sigma$ limiting magnitude per pointing in this 
survey is 20.6 \citep{ZTF2019}. As a Pan-STARRS-based survey example, we consider a strategy which is alternative to ZTF. 
The sky coverage is only 20 per cent of that in ZTF, but in $griz$ bands with the same cadence as ZTF. All lensed supernovae detectable 
in this survey would be found in the $i$ band for which the survey would reach a $5\sigma$ limiting magnitude of 20.6 per pointing. 
The potentially more effective $z$ band would not improve detection rates due to its shallower depth of 20.2. Both ZTF and 
the Pan-STARRS survey operate in the regime of limiting magnitudes at which the only effective detection technique is the 
magnification method. Therefore, we omit the rates expected for the image multiplicity and hybrid techniques for them in Table~\ref{rates_survey}.

\begin{table*}
\begin{tabular}{lccrrrrr}

survey/detection method & effective area & Type Ia & Type IIP & Type IIL & Type Ib/c & Type IIn \\
\hline
ZTF/magnification & 15 000 deg$^{2}$ & 2.1 (\textit{r}) & 0.37 (\textit{r}) & 0.23 (\textit{r}) & 0.36 (\textit{r}) & 3.8 (\textit{r}) \\
Pan-STARRS/magnification & 3000 deg$^{2}$ &  0.9 (\textit{i})& 0.20 (\textit{i})& 0.13 (\textit{i}) & 0.20 (\textit{i}) & 0.8 (\textit{i})\\
LSST/magnification & 20 000 deg$^{2}$       &  61(\textit{z}) & 12.2 (\textit{z}) & 8.7 (\textit{y}) & 12.3 (\textit{y})& 184 (\textit{g})\\
LSST/image multiplicity & 20 000 deg$^{2}$ & 44 (\textit{i}) & 6.1 (\textit{i}) & 5.5 (\textit{i}) & 6.8 (\textit{i}) & 88 (\textit{g})\\
LSST/hybrid & 20 000 deg$^{2}$                   &  89 (\textit{iz}) &   16.3 (\textit{iz}) &  11.9 (\textit{iy}) & 15.8 (\textit{iy}) & 210 (\textit{g})\\
\end{tabular}
\caption{The expected \textit{annual} numbers of discovered lensed supernovae in ongoing or upcoming transient surveys. 
The method based on image multiplicity is ineffective in case case of ZTF and Pan-STARRS, and thus it is omitted in the table. Symbols 
in parentheses indicate the most effective filter yielding the largest number of detections.
}
\label{rates_survey}
\end{table*}

LSST will cover about 20000~deg$^{2}$ in 6 $ugrizy$ bands every 2--3 weeks with a limiting magnitude of 
24 \citep{LSST2009,LSST2017}. Precise estimation of discovery rates relies on details of the observational strategy which 
is yet to be decided.  However, reasonable estimates can be obtained based on the following reasoning. A typical redshift of 
strongly lensed supernovae to be discovered by LSST is $z=1$. This means that the faintest supernovae can be detected only 
within a time window of about 10 days (rest frame) around the peak of their light curve (20 days in the observer frame). 
In order to satisfy this condition, all detectable supernovae should be at least 0.2 mag brighter at the peak than the 
actual limiting magnitudes of the survey which are $(24.5,\,24.2,\,23.6,\,22.8,\,22.0)$ in the $grizy$ bands 
(mean $5\sigma$ per pointing; based on LSST collaboration's simulations, baseline2018a 
run\footnote{https://www.lsst.org/scientists/simulations/opsim/opsim-survey-data}, which is the current official 
reference simulated survey). Then, the expected discovery rate can be found as the maximum rate found for all filters. 
Since LSST will observe down to limiting magnitudes at which the image multiplicity method becomes effective, we provide 
rate estimates for all three methods of finding lensed supernova candidates. Table~\ref{rates_survey} also shows the most 
effective filter (with the highest rate) for different supernova types and detection methods.

Pre-LSST surveys may find about 5 strongly lensed supernovae per year, based solely on the magnification method. 
The most frequent type will be IIn, followed by Ia. Discovery rates will increase by nearly two orders of magnitude for 
LSST. It is also clear that the magnification method is expected to yield only about 2 times more discoveries than the 
image multiplication method. However, both methods will detect lensed supernovae at comparable rates when we assume that 
the measured flux in the magnification methods comes solely from the brightest image (see Fig.~\ref{yields_comparison}). 
Furthermore, keeping in mind that multiply imaged transients do not require follow-up observations confirming their lensing 
nature (in contrast to highly magnified transients which can be confused with superluminous supernovae) we conclude 
that \textit{the image multiplicity method will be the most effective technique for finding lensed transients in LSST data.}

Lensed supernovae discovered by LSST will be typically detected in the $i$ band as a multiply imaged transient and in the 
$y$ band (or the $z$ band for type Ia and IIP) as a highly magnified transient (except for type IIn which will be discovered 
primarily in bluer filters). The difference in the effective discovery band can strengthen the complementarity of the two detection techniques. Considering a hybrid method of finding lensed supernovae via image multiplicity in the $i$ band and 
magnification in the $y$ band (or $z$ for Type Ia and IIP), we find that discovery rates in this approach are higher 
by 30--50 per cent than those based solely on magnification. 

Except for type IIP supernovae, our discovery rates estimated for ZTF and LSST based on the magnification method agree 
fairly well with analogous predictions obtained by \citet{Gold2018}. 
Adopting the same detection conditions and survey parameters for LSST, we also recover fairly closely the rates estimated by 
\citet{Ogu2010a}, with the total number of lensed Type Ia and core-collapse supernovae to be discovered by LSST of 32 and 36. 
These rates are lower by a factor of $>10$ than those listed in Table~\ref{yields_comparison} and obtained by \citet{Gold2018}. 
These differences can be fully accounted for
to the stricter detection criteria adopted by \citet{Ogu2010a}, i.e.\ an effective limiting magnitude of 22.6 in the 
$i$ band (in order to sample light curves at minimum depth of 0.7 mag around the peak) and an effective survey time of 
$2.5$ years (accounting for seasonal changes of the surveyed area), and do not signal discrepancies in the basic
methodologies in the independent approaches to estimating detection rates. It is worth 
mentioning that more restrictive selection criteria than those proposed by \citet{Ogu2010a} are required for obtaining high 
quality measurements of gravitational time delays. In particular, a minimum precision of 5 per cent and an accuracy of 
1 per cent in the time delay measurements (if based solely on LSST observations) would reduce the number of 
lensed type Ia supernovae to about 1 per year \citep{Hub2019}. This rate can be increased by a factor of 2--16 by 
employing other instruments for follow-up observations.

\subsection{Observed fractions of supernova types}

\begin{figure}
\centering
\includegraphics[width=0.48\textwidth]{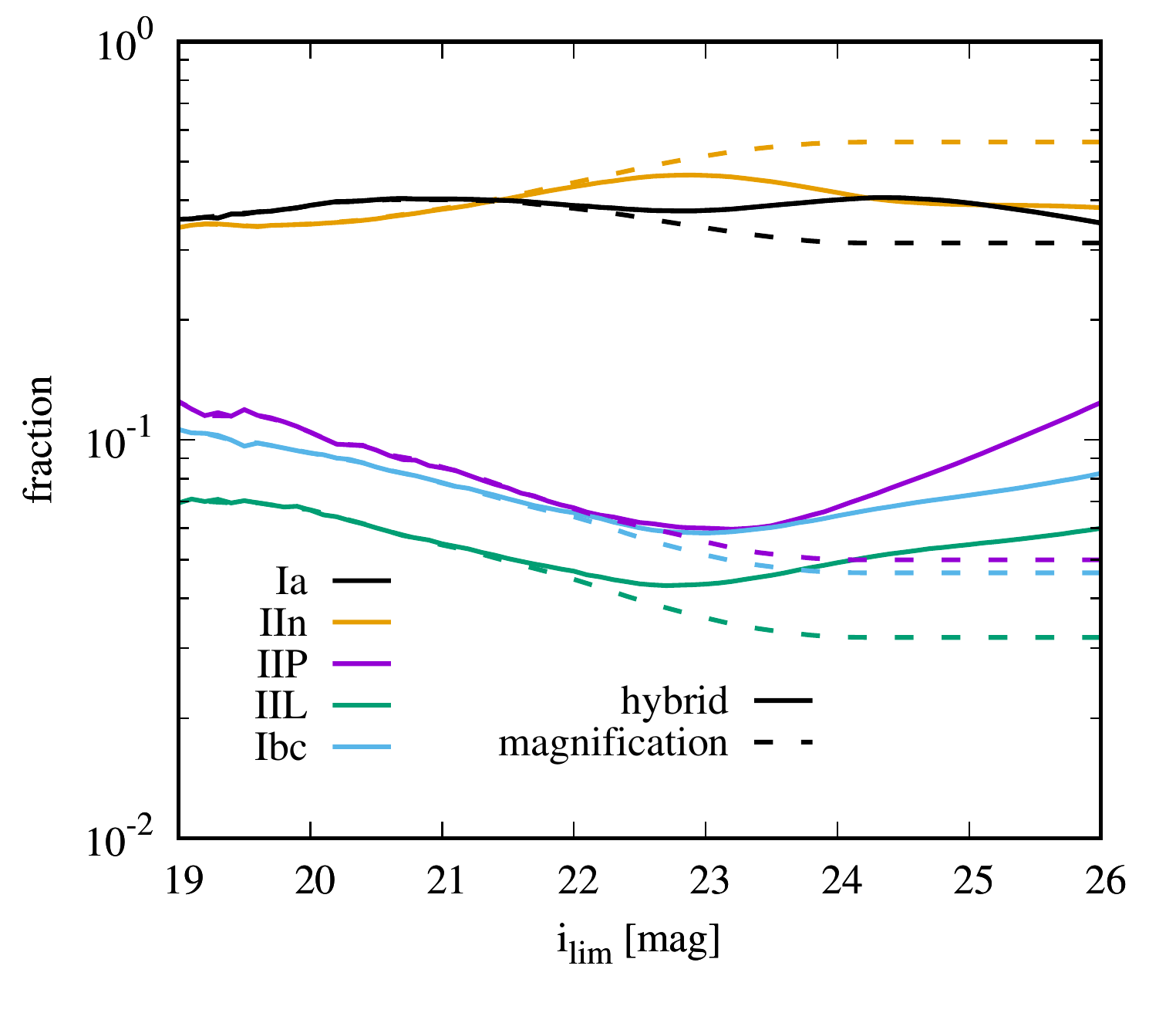}
\caption{Fractions of different supernova types in samples of gravitationally lensed supernovae detected in transient 
surveys with limiting magnitude $i_{\rm lim}$ in the $i$ band. Gravitationally lensed supernovae are detected via magnification 
(dashed curves) or using the hybrid method (solid curves).
}
\label{fractions}
\end{figure}

Fig.~\ref{fractions} shows predicted fractions of different types of supernovae in representative samples of  gravitationally 
lensed supernovae expected in transient surveys with a range of limiting magnitudes between 19 and 26. 

The observed fractions depend quite weakly on limiting magnitude. The most noticeable trend occurs for type IIP, IIL and Ibc at large 
limiting magnitudes for the image multiplicity method. The highest fractions of these supernovae are expected for extremely shallow or 
deep surveys.

Type Ia and IIn supernovae clearly dominate detections, with fractions of about 30 per cent each. 
The prevalence of type IIn supernovae becomes even stronger Ia when one 
includes detections in the $g$ band, the most efficient filter for observing type IIn supernovae. We emphasize that the high 
relative discovery rates of type IIn supernovae rely quite strongly on the adopted spectral templates which in turn depend on 
the extinction correction performed in the analysis of the observational data used here \citep[see][]{DiC2002}.

Nevertheless, the high predicted relative detection rates of lensed type IIn supernovae is intriguing. The rates and luminosity distribution of type IIn supernovae at high redshift are quite uncertain, theoretically because of their wide range of possible
progenitors, including very massive stars in the tail of the initial mass function \citep[e.g.,][]{2009Natur.458..865G, 2014ARA&A..52..487S, 2017A&A...599A.129T} and 
observationally, because of their diverse photometric properties \citep[e.g.,][]{2013A&A...555A..10T, 2014AJ....147..118R, 2015A&A...584A..62C}. Fortunately, lensed type IIn supernovae can readily be distinguished from other supernovae due to their 
narrow (and hence high spectroscopic signal-to-noise ratio) Balmer emission lines. Hence, observations of lensed IIn could provide unique insights into their intrinsic properties at high redshift.

\subsection{Discovery space}

\begin{figure*}
\centering
\includegraphics[trim={0.cm 0.cm 0.cm 0.0cm},clip,width=1.\textwidth]{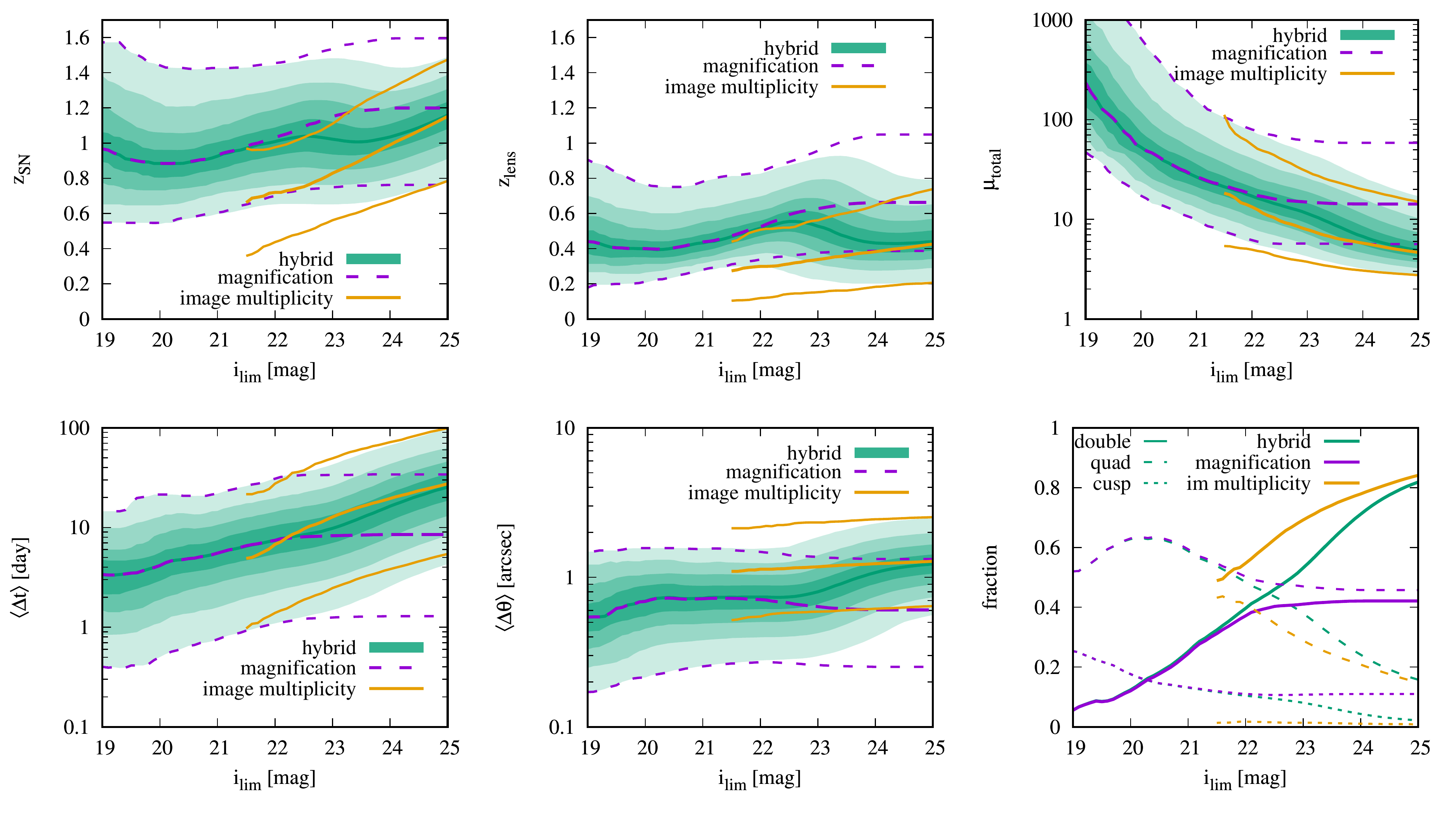}
\caption{Discovery space of gravitationally lensed type Ia supernovae detected with different methods in a single-filter survey 
with depth $i_{\rm lim}$ in the $i$ band. The panels show distributions of supernova redshift $z_{\rm SN}$, 
lens redshift $z_{\rm lens}$, 
total magnification $\mu_{\rm total}$, gravitational time delay averaged over all image pairs $\langle \Delta t\rangle$ and 
image separation averaged over all image pairs $\langle\Delta \theta \rangle$. The green contours show 10-quantiles of 
the distributions for supernova samples produced by the hybrid method. The purple and orange curves show the median and 
$80$-percent probability range of probability distributions for the magnification and image multiplicity methods, respectively. 
The results for the image multiplicity are shown only for detection rates larger than 10 per cent of the corresponding 
detection rates for the magnification method. The bottom right panel shows fractions of different lensing 
configurations with two images (doubles), three images (cusps) or four images (quads).
}
\label{phase-space}
\end{figure*}

The detection methods considered in this study may also lead to differences in the phase space of lensing configurations. 
Fig.~\ref{phase-space} shows the distributions of basic lensing parameters of lensed type Ia supernovae as a function of limiting 
$i$-band magnitude. The green contours shows 10-quantiles of the distributions expected for the hybrid method. The purple and orange 
curves show the median and a range containing 80 per cent of the cases for the magnification and image multiplicity techniques, 
respectively. We omit results for the image multiplication technique at limiting magnitudes lower than $i_{\rm lim}=21.5$ where the 
method is far less competitive, with the potential discovery rate smaller than 10 per cent of that of the magnification technique. 
Fig.~\ref{phase-space} also shows the fractions of the different image configurations with two (doubles), three (cusps) and 
four images (quads).

Gravitationally lensed supernovae to be found in ongoing pre-LSST shallow surveys will be extremely magnified. 
For a limiting magnitude of 20.6 (depth of ZTF), the expected mean magnification is 20 and a 10 per cent tail of the 
distribution includes cases with magnifications exceeding 100. For shallower surveys, the expected magnification become 
substantially higher with the mean reaching $\mu=100$ at $i_{\rm lim}=19.5$. 

The expected lensing properties of lensed supernovae detected via magnification pose a challenge to using them as cosmological 
probes. In particular, the typical time delays for lensed supernovae found in pre-LSST surveys are below 10 days and typical 
image separations will hardly exceed the arcsec scale. Fig.~\ref{phase-space} demonstrates that LSST or deeper surveys -- 
with typical time delays always below 10 days -- can hardly mitigate this problem. In this respect, the image multiplicity method 
appears to be more promising. When applied to LSST-like surveys ($i_{\rm lim}=23.5$), the method is expected to find lensed 
supernovae with the mean time delay of 20 days (and a 10 per cent tail of the distribution with time delays at least 60 days) 
and typical image separations larger than 1 arcsec. Undoubtedly, this increases the potential for using lensed supernovae to 
place robust cosmological constraints.

Time delays and image separations are not the only differences between the populations of lensed supernovae found via magnification 
and image multiplicity. We also find that the image multiplicity method is more sensitive to supernovae and lens galaxies at lower 
redshifts. There is also a clear difference in terms of image configuration. A much larger fraction of lensed supernovae 
found via image multiplicity are doubles, whereas for the magnification methods the numbers of quads and doubles are comparable. 
Unsurprisingly, both methods exhibit an increasing (decreasing) trend in the fraction of doubles (quads) with increasing limiting 
magnitude. The fraction of cups is at the sub-percent level for the image magnification methods, whereas it reaches a 10-precent level 
for the magnification  technique.

\begin{figure}
\centering
\includegraphics[width=0.48\textwidth]{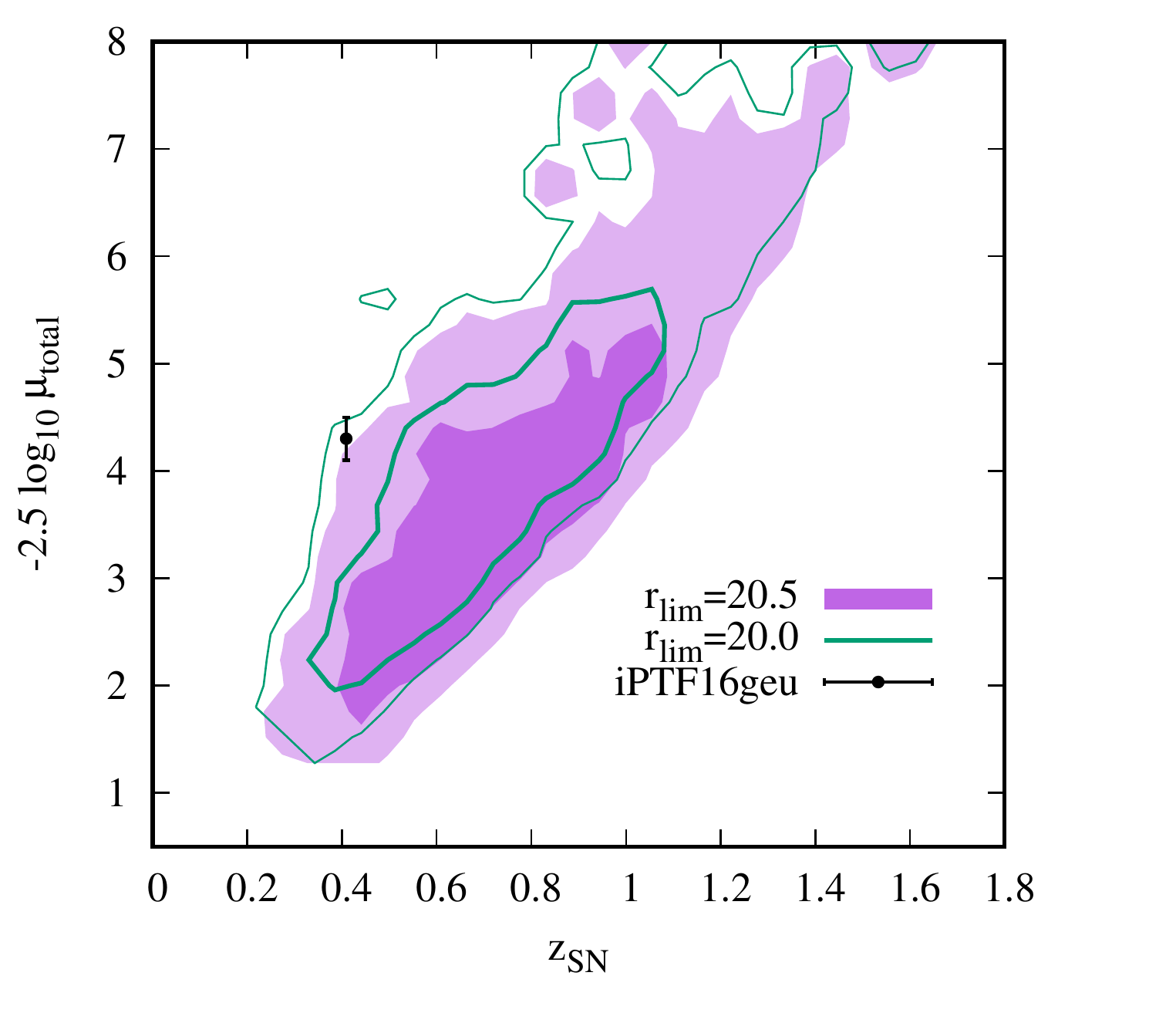}
\caption{The expected distribution of the total magnification and redshift (in the form of $1\sigma$ and $2\sigma$ credibility contours) of lensed supernovae detectable in the intermediate Palomar Transient Factory (iPTF) survey with limiting magnitude 
$r_{\rm lim}$ in the $r$ band, compared to the measured parameters of observed gravitationally lensed supernovae iPTF16geu. 
The observed supernovae lies well within a $2\sigma$ contour.
}
\label{iPTF}
\end{figure}

In the context of discussing the future phase space of lensing configurations, it is interesting to consider the case of the 
gravitationally lensed supernova iPTF16geu. The supernova was discovered in a relatively shallow survey. In this respect, its 
high magnification of $\mu_{\rm total}\sim 50$ is not surprising. However, the observed magnification turns out to be at odds with 
the theoretical expectations when the redshift of the supernova is taken into account \citep{2017Mor,Gold2018}. In order to address 
this tension, we compute two-dimensional credibility contours for redshifts and gravitational magnifications of all lensed supernovae 
detectable in a transient survey equivalent to the intermediate Palomar Transient Factory. As a nominal depth of the survey, we 
consider either 20.5 or 20.0 in the $r$ band. The latter is a more appropriate choice when one requires good observations of light 
curves around the peak \citep{2017Mor}. The supernova was discovered as a peculiarly bright, initially unresolved, transient; 
therefore, it is justified to employ the magnification method as an effective approach to finding lensed transients in the 
iPTF survey. As shown in Fig.~\ref{iPTF}, the tension between iPTF16geu and theoretical predictions is at a quite modest level 
of $2\sigma$. We conclude that the peculiar lensing configuration of iPTF16geu can be simply a statistical fluke. However, 
the flux ratio anomalies between observed brightness of the images and the best fit lens model pose a serious problem because 
the discrepancy seems too large to be ascribed to microlensing \citep{Yah2017}.

\section{Summary and conclusions}

We have compared different observational strategies for detecting gravitationally lensed supernovae in massive transient surveys. 
The strategies rely on finding multiply imaged transients or highly magnified supernovae (deduced via comparing observed magnitudes 
to fiducial magnitudes of a type Ia supernova located in the apparent host galaxy). Adopting state-of-the-art models of lens galaxies 
constrained by the SDSS data and the standard cosmological $\Lambda$CDM mode, we have calculated detection rates for each method in 
5 \textit{grizy} bands for the main supernova classes including: type Ia, core-collapse supernovae types IIP, IIL, Ibc and IIn. 
We provide simple fitting functions approximating the computed rates.

We find that detecting lensed supernovae as strongly magnified transients is the only effective detection method working for shallow 
pre-LSST surveys with limiting magnitudes smaller than 22. However, the expected yields saturate at limiting magnitudes of about 
23.0--23.5 (depending on supernova type and filter) where lensed supernovae are more likely to be fainter than a fiducial type Ia 
supernova in the apparent host galaxies (lens galaxy). At this limiting magnitude, the magnification and image multiplicity method 
yield comparable numbers of lensed supernovae. Supernovae found by the two methods are to a large extent independent; therefore, 
a noticeable improvement of discovery rates can be achieved by combining the two methods. The resulting hybrid method increases the 
yields by 50 per cent at limiting magnitudes corresponding to comparable rates expected for the two primary methods. For larger 
limiting magnitudes $\gtrsim23.5$, the image multiplicity method completely surpasses the performance of the magnification technique.

Detection rates depend strongly on filter. Except for type IIn, discovery rates decrease with decreasing effective wavelength. An inverse 
trend found for type IIn supernovae results from the presence of strong UV flux and a relatively higher incidence of luminous supernovae 
with $M_{B}<-20$. In the overall counts of gravitationally lensed supernovae, type IIn will be the most common class followed by type Ia,
regardless of the adopted detection method. The rates are dominated by intrinsically bright supernovae; therefore, the robustness of our 
predictions rely on accurate modelling of high-luminosity tails of the supernova luminosity functions, currently approximated by 
Gaussian distributions.

Revisiting the initial comparison between the image multiplicity and magnification methods made by \citet{Gold2017}, 
we find that detecting lensed supernovae via image multiplicity is as efficient as the magnification method at the limiting magnitudes 
of LSST and it surpasses the latter for deeper surveys. Moreover, strongly lensed supernovae found via image multiplicity are also 
characterized by longer time delays and larger image separations than systems found via magnification. This makes the image multiplicity 
method a more appealing observational strategy for LSST when considering the potential of using lensed supernovae as cosmological probes. 
It is also worth mentioning that the image multiplicity approach does not rely on follow-up observations and it can be naturally applied 
to self-contained massive surveys without auxiliary imaging observations. In contrast, candidates for lensed supernovae selected via 
magnification naturally require follow-up observations to confirm their the lensing nature. 

We estimate that ZTF with a depth of 20.6 in the $r$ band will detect 1.9 type Ia 4.1 core-collapse (primarily IIn) lensed supernovae 
per year. Analogous computations for LSST yield 44 type Ia and 106 core-collapse (primarily IIn) lensed supernovae per year, detected 
via image multiplicity. The hybrid method will allow to increase these rates to 89 and 254 detections per year for type Ia and 
core-collapse, respectively. Core-collapse lensed supernovae will be dominated by type IIn with relative fraction of 80 per cent. 

\section*{Acknowledgments}
We are highly indebted to the referee, Prasenjit Saha, for his insightful comments and helpful suggestions. This work was supported 
by a VILLUM FONDEN Investigator grant to JH (project number 16599).

\bibliography{master}

\begin{thebibliography}{78}
\expandafter\ifx\csname natexlab\endcsname\relax\def\natexlab#1{#1}\fi

\bibitem[{{Andersen} \& {Hjorth}(2018)}]{And2018}
{Andersen} P., {Hjorth} J., 2018, \mnras, 480, 68

\bibitem[{{Anderson} {et~al}\mbox{.}(2014){Anderson},
  {Gonz{\'a}lez-Gait{\'a}n}, {Hamuy}, {Guti{\'e}rrez}, {Stritzinger}, {Olivares
  E.}, {Phillips}, {Schulze}, {Antezana}, {Bolt}, {Campillay}, {Castell{\'o}n},
  {Contreras}, {de Jaeger}, {Folatelli}, {F{\"o}rster}, {Freedman},
  {Gonz{\'a}lez}, {Hsiao}, {Krzemi{\'n}ski}, {Krisciunas}, {Maza}, {McCarthy},
  {Morrell}, {Persson}, {Roth}, {Salgado}, {Suntzeff}, \&
  {Thomas-Osip}}]{2014ApJ...786...67A}
{Anderson} J.~P. {et~al.}, 2014, \apj, 786, 67

\bibitem[{{Barbary} {et~al}\mbox{.}(2012){Barbary}, {Aldering}, {Amanullah},
  {Brodwin}, {Connolly}, {Dawson}, {Doi}, {Eisenhardt}, {Faccioli}, {Fadeyev},
  {Fakhouri}, {Fruchter}, {Gilbank}, {Gladders}, {Goldhaber}, {Goobar},
  {Hattori}, {Hsiao}, {Huang}, {Ihara}, {Kashikawa}, {Koester}, {Konishi},
  {Kowalski}, {Lidman}, {Lubin}, {Meyers}, {Morokuma}, {Oda}, {Panagia},
  {Perlmutter}, {Postman}, {Ripoche}, {Rosati}, {Rubin}, {Schlegel},
  {Spadafora}, {Stanford}, {Strovink}, {Suzuki}, {Takanashi}, {Tokita},
  {Yasuda}, \& {Supernova Cosmology Project}}]{Bar2012}
{Barbary} K. {et~al.}, 2012, \apj, 745, 31

\bibitem[{{Belczynski} {et~al}\mbox{.}(2005){Belczynski}, {Bulik}, \&
  {Ruiter}}]{Bel2005}
{Belczynski} K., {Bulik} T., {Ruiter} A.~J., 2005, \apj, 629, 915

\bibitem[{{Bellm} {et~al}\mbox{.}(2019){Bellm}, {Kulkarni}, {Graham}, {Dekany},
  {Smith}, {Riddle}, {Masci}, {Helou}, {Prince}, {Adams}, {Barbarino},
  {Barlow}, {Bauer}, {Beck}, {Belicki}, {Biswas}, {Blagorodnova}, {Bodewits},
  {Bolin}, {Brinnel}, {Brooke}, {Bue}, {Bulla}, {Burruss}, {Cenko}, {Chang},
  {Connolly}, {Coughlin}, {Cromer}, {Cunningham}, {De}, {Delacroix}, {Desai},
  {Duev}, {Eadie}, {Farnham}, {Feeney}, {Feindt}, {Flynn}, {Franckowiak},
  {Frederick}, {Fremling}, {Gal-Yam}, {Gezari}, {Giomi}, {Goldstein},
  {Golkhou}, {Goobar}, {Groom}, {Hacopians}, {Hale}, {Henning}, {Ho}, {Hover},
  {Howell}, {Hung}, {Huppenkothen}, {Imel}, {Ip}, {Ivezi{\'c}}, {Jackson},
  {Jones}, {Juric}, {Kasliwal}, {Kaspi}, {Kaye}, {Kelley}, {Kowalski},
  {Kramer}, {Kupfer}, {Landry}, {Laher}, {Lee}, {Lin}, {Lin}, {Lunnan},
  {Giomi}, {Mahabal}, {Mao}, {Miller}, {Monkewitz}, {Murphy}, {Ngeow},
  {Nordin}, {Nugent}, {Ofek}, {Patterson}, {Penprase}, {Porter}, {Rauch},
  {Rebbapragada}, {Reiley}, {Rigault}, {Rodriguez}, {van Roestel}, {Rusholme},
  {van Santen}, {Schulze}, {Shupe}, {Singer}, {Soumagnac}, {Stein}, {Surace},
  {Sollerman}, {Szkody}, {Taddia}, {Terek}, {Van Sistine}, {van Velzen},
  {Vestrand}, {Walters}, {Ward}, {Ye}, {Yu}, {Yan}, \& {Zolkower}}]{ZTF2019}
{Bellm} E.~C. {et~al.}, 2019, Publications of the Astronomical Society of the
  Pacific, 131, 018002

\bibitem[{{Bezanson} {et~al}\mbox{.}(2011){Bezanson}, {van Dokkum}, {Franx},
  {Brammer}, {Brinchmann}, {Kriek}, {Labb{\'e}}, {Quadri}, {Rix}, {van de
  Sande}, {Whitaker}, \& {Williams}}]{Bez2011}
{Bezanson} R. {et~al.}, 2011, \apj, 737, L31

\bibitem[{{Blanc} {et~al}\mbox{.}(2004){Blanc}, {Afonso}, {Alard}, {Albert},
  {Aldering}, {Amadon}, {Andersen}, {Ansari}, {Aubourg}, {Balland}, {Bareyre},
  {Beaulieu}, {Charlot}, {Conley}, {Coutures}, {Dahl{\'e}n}, {Derue}, {Fan},
  {Ferlet}, {Folatelli}, {Fouqu{\'e}}, {Garavini}, {Glicenstein}, {Goldman},
  {Goobar}, {Gould}, {Graff}, {Gros}, {Haissinski}, {Hamadache}, {Hardin},
  {Hook}, {de Kat}, {Kent}, {Kim}, {Lasserre}, {Le Guillou}, {Lesquoy}, {Loup},
  {Magneville}, {Marquette}, {Maurice}, {Maury}, {Milsztajn}, {Moniez},
  {Mouchet}, {Newberg}, {Nobili}, {Palanque-Delabrouille}, {Perdereau},
  {Pr{\'e}vot}, {Rahal}, {Regnault}, {Rich}, {Ruiz-Lapuente}, {Spiro},
  {Tisserand}, {Vidal-Madjar}, {Vigroux}, {Walton}, \& {Zylberajch}}]{Bla2004}
{Blanc} G. {et~al.}, 2004, \aap, 423, 881

\bibitem[{{Botticella} {et~al}\mbox{.}(2008){Botticella}, {Riello},
  {Cappellaro}, {Benetti}, {Altavilla}, {Pastorello}, {Turatto}, {Greggio},
  {Patat}, {Valenti}, {Zampieri}, {Harutyunyan}, {Pignata}, \&
  {Taubenberger}}]{Bot2008}
{Botticella} M.~T. {et~al.}, 2008, \aap, 479, 49

\bibitem[{{Cano} {et~al}\mbox{.}(2018){Cano}, {Selsing}, {Hjorth}, {de Ugarte
  Postigo}, {Christensen}, {Gall}, \& {Kann}}]{Can2018}
{Cano} Z., {Selsing} J., {Hjorth} J., {de Ugarte Postigo} A., {Christensen} L.,
  {Gall} C., {Kann} D.~A., 2018, \mnras, 473, 4257

\bibitem[{{Cappellaro} {et~al}\mbox{.}(2015){Cappellaro}, {Botticella},
  {Pignata}, {Grado}, {Greggio}, {Limatola}, {Vaccari}, {Baruffolo}, {Benetti},
  {Bufano}, {Capaccioli}, {Cascone}, {Covone}, {De Cicco}, {Falocco}, {Della
  Valle}, {Jarvis}, {Marchetti}, {Napolitano}, {Paolillo}, {Pastorello},
  {Radovich}, {Schipani}, {Spiro}, {Tomasella}, \&
  {Turatto}}]{2015A&A...584A..62C}
{Cappellaro} E. {et~al.}, 2015, \aap, 584, A62

\bibitem[{{Cappellaro} {et~al}\mbox{.}(1999){Cappellaro}, {Evans}, \&
  {Turatto}}]{Cap1999}
{Cappellaro} E., {Evans} R., {Turatto} M., 1999, \aap, 351, 459

\bibitem[{{Castro} {et~al}\mbox{.}(2018){Castro}, {Quartin}, {Giocoli},
  {Borgani}, \& {Dolag}}]{Cas2018}
{Castro} T., {Quartin} M., {Giocoli} C., {Borgani} S., {Dolag} K., 2018,
  \mnras, 478, 1305

\bibitem[{{Chae}(2003)}]{Chae2003}
{Chae} K.-H., 2003, \mnras, 346, 746

\bibitem[{{Chen} {et~al}\mbox{.}(2019){Chen}, {Kelly}, {Diego}, {Oguri},
  {Williams}, {Zitrin}, {Treu}, {Smith}, {Broadhurst}, {Kaiser}, {Foley},
  {Filippenko}, {Salo}, {Hjorth}, \& {Selsing}}]{Chen2019}
{Chen} W. {et~al.}, 2019, arXiv e-prints, arXiv:1902.05510

\bibitem[{{Choi} {et~al}\mbox{.}(2007){Choi}, {Park}, \& {Vogeley}}]{Choi2007}
{Choi} Y.-Y., {Park} C., {Vogeley} M.~S., 2007, \apj, 658, 884

\bibitem[{{Dahlen} {et~al}\mbox{.}(2008){Dahlen}, {Strolger}, \&
  {Riess}}]{Dah2008}
{Dahlen} T., {Strolger} L.-G., {Riess} A.~G., 2008, \apj, 681, 462

\bibitem[{{Di Carlo} {et~al}\mbox{.}(2002){Di Carlo}, {Massi}, {Valentini}, {Di
  Paola}, {D'Alessio}, {Brocato}, {Guidubaldi}, {Dolci}, {Pedichini},
  {Speziali}, {Li Causi}, {Caratti o Garatti}, {Cappellaro}, {Turatto},
  {Arkharov}, {Gnedin}, {Larionov}, {Benetti}, {Pastorello}, {Aretxaga},
  {Chavushyan}, {Vega}, {Danziger}, \& {Tornamb{\'e}}}]{DiC2002}
{Di Carlo} E. {et~al.}, 2002, \apj, 573, 144

\bibitem[{{Diego}(2018)}]{Diego2018}
{Diego} J.~M., 2018, arXiv e-prints, arXiv:1806.04668

\bibitem[{{Dilday} {et~al}\mbox{.}(2008){Dilday}, {Kessler}, {Frieman},
  {Holtzman}, {Marriner}, {Miknaitis}, {Nichol}, {Romani}, {Sako}, {Bassett},
  {Becker}, {Cinabro}, {DeJongh}, {Depoy}, {Doi}, {Garnavich}, {Hogan}, {Jha},
  {Konishi}, {Lampeitl}, {Marshall}, {McGinnis}, {Prieto}, {Riess}, {Richmond},
  {Schneider}, {Smith}, {Takanashi}, {Tokita}, {van der Heyden}, {Yasuda},
  {Zheng}, {Barentine}, {Brewington}, {Choi}, {Crotts}, {Dembicky}, {Harvanek},
  {Im}, {Ketzeback}, {Kleinman}, {Krzesi{\'n}ski}, {Long}, {Malanushenko},
  {Malanushenko}, {McMillan}, {Nitta}, {Pan}, {Saurage}, {Snedden}, {Watters},
  {Wheeler}, \& {York}}]{Dil2008}
{Dilday} B. {et~al.}, 2008, \apj, 682, 262

\bibitem[{{Dilday} {et~al}\mbox{.}(2010){Dilday}, {Smith}, {Bassett}, {Becker},
  {Bender}, {Castander}, {Cinabro}, {Filippenko}, {Frieman}, {Galbany},
  {Garnavich}, {Goobar}, {Hopp}, {Ihara}, {Jha}, {Kessler}, {Lampeitl},
  {Marriner}, {Miquel}, {Moll{\'a}}, {Nichol}, {Nordin}, {Riess}, {Sako},
  {Schneider}, {Sollerman}, {Wheeler}, {{\"O}stman}, {Bizyaev}, {Brewington},
  {Malanushenko}, {Malanushenko}, {Oravetz}, {Pan}, {Simmons}, \&
  {Snedden}}]{Dil2010}
{Dilday} B. {et~al.}, 2010, \apj, 713, 1026

\bibitem[{{Gal-Yam} \& {Leonard}(2009)}]{2009Natur.458..865G}
{Gal-Yam} A., {Leonard} D.~C., 2009, \nat, 458, 865

\bibitem[{{Gilliland} {et~al}\mbox{.}(1999){Gilliland}, {Nugent}, \&
  {Phillips}}]{Gil1999}
{Gilliland} R.~L., {Nugent} P.~E., {Phillips} M.~M., 1999, \apj, 521, 30

\bibitem[{{Goldstein} \& {Nugent}(2017)}]{Gold2017}
{Goldstein} D.~A., {Nugent} P.~E., 2017, \apjl, 834, L5

\bibitem[{{Goldstein} {et~al}\mbox{.}(2018{\natexlab{a}}){Goldstein}, {Nugent},
  \& {Goobar}}]{Gold2018}
{Goldstein} D.~A., {Nugent} P.~E., {Goobar} A., 2018{\natexlab{a}}, ArXiv
  e-prints, arXiv:1809.10147

\bibitem[{{Goldstein} {et~al}\mbox{.}(2018{\natexlab{b}}){Goldstein}, {Nugent},
  {Kasen}, \& {Collett}}]{Gold2018a}
{Goldstein} D.~A., {Nugent} P.~E., {Kasen} D.~N., {Collett} T.~E.,
  2018{\natexlab{b}}, \apj, 855, 22

\bibitem[{{Goobar} {et~al}\mbox{.}(2017){Goobar}, {Amanullah}, {Kulkarni},
  {Nugent}, {Johansson}, {Steidel}, {Law}, {M{\"o}rtsell}, {Quimby},
  {Blagorodnova}, {Brandeker}, {Cao}, {Cooray}, {Ferretti}, {Fremling},
  {Hangard}, {Kasliwal}, {Kupfer}, {Lunnan}, {Masci}, {Miller}, {Nayyeri},
  {Neill}, {Ofek}, {Papadogiannakis}, {Petrushevska}, {Ravi}, {Sollerman},
  {Sullivan}, {Taddia}, {Walters}, {Wilson}, {Yan}, \& {Yaron}}]{Goo2017}
{Goobar} A. {et~al.}, 2017, Science, 356, 291

\bibitem[{{Graur} {et~al}\mbox{.}(2017){Graur}, {Bianco}, {Modjaz}, {Shivvers},
  {Filippenko}, {Li}, \& {Smith}}]{Gra2017}
{Graur} O., {Bianco} F.~B., {Modjaz} M., {Shivvers} I., {Filippenko} A.~V.,
  {Li} W., {Smith} N., 2017, \apj, 837, 121

\bibitem[{{Graur} \& {Maoz}(2013)}]{Gra2013}
{Graur} O., {Maoz} D., 2013, \mnras, 430, 1746

\bibitem[{{Graur} {et~al}\mbox{.}(2011){Graur}, {Poznanski}, {Maoz}, {Yasuda},
  {Totani}, {Fukugita}, {Filippenko}, {Foley}, {Silverman}, {Gal-Yam},
  {Horesh}, \& {Jannuzi}}]{Gra2011}
{Graur} O. {et~al.}, 2011, \mnras, 417, 916

\bibitem[{{Graur} {et~al}\mbox{.}(2014){Graur}, {Rodney}, {Maoz}, {Riess},
  {Jha}, {Postman}, {Dahlen}, {Holoien}, {McCully}, {Patel}, {Strolger},
  {Ben{\'{\i}}tez}, {Coe}, {Jouvel}, {Medezinski}, {Molino}, {Nonino},
  {Bradley}, {Koekemoer}, {Balestra}, {Cenko}, {Clubb}, {Dickinson},
  {Filippenko}, {Frederiksen}, {Garnavich}, {Hjorth}, {Jones}, {Leibundgut},
  {Matheson}, {Mobasher}, {Rosati}, {Silverman}, {U}, {Jedruszczuk}, {Li},
  {Lin}, {Mirmelstein}, {Neustadt}, {Ovadia}, \& {Rogers}}]{Gra2014}
{Graur} O. {et~al.}, 2014, \apj, 783, 28

\bibitem[{{Grillo} {et~al}\mbox{.}(2018){Grillo}, {Rosati}, {Suyu}, {Balestra},
  {Caminha}, {Halkola}, {Kelly}, {Lombardi}, {Mercurio}, {Rodney}, \&
  {Treu}}]{Gri2018}
{Grillo} C. {et~al.}, 2018, \apj, 860, 94

\bibitem[{{Hardin} {et~al}\mbox{.}(2000){Hardin}, {Afonso}, {Alard}, {Albert},
  {Amadon}, {Andersen}, {Ansari}, {Aubourg}, {Bareyre}, {Bauer}, {Beaulieu},
  {Blanc}, {Bouquet}, {Char}, {Charlot}, {Couchot}, {Coutures}, {Derue},
  {Ferlet}, {Glicenstein}, {Goldman}, {Gould}, {Graff}, {Gros}, {Haissinski},
  {Hamilton}, {de Kat}, {Kim}, {Lasserre}, {Lesquoy}, {Loup}, {Magneville},
  {Mansoux}, {Marquette}, {Maurice}, {Milsztajn}, {Moniez},
  {Palanque-Delabrouille}, {Perdereau}, {Pr{\'e}vot}, {Regnault}, {Rich},
  {Spiro}, {Vidal-Madjar}, {Vigroux}, {Zylberajch}, \& {EROS
  Collaboration}}]{Har2000}
{Hardin} D. {et~al.}, 2000, \aap, 362, 419

\bibitem[{{Holder} \& {Schechter}(2003)}]{Hol2003}
{Holder} G.~P., {Schechter} P.~L., 2003, \apj, 589, 688

\bibitem[{{Horesh} {et~al}\mbox{.}(2008){Horesh}, {Poznanski}, {Ofek}, \&
  {Maoz}}]{Hor2008}
{Horesh} A., {Poznanski} D., {Ofek} E.~O., {Maoz} D., 2008, \mnras, 389, 1871

\bibitem[{{Huber} {et~al}\mbox{.}(2019){Huber}, {Suyu}, {Noebauer}, {Bonvin},
  {Rothchild}, {Chan}, {Awan}, {Courbin}, {Kromer}, {Marshall}, {Oguri},
  {Ribeiro}, \& {The LSST Dark Energy Science Collaboration}}]{Hub2019}
{Huber} S. {et~al.}, 2019, arXiv e-prints, arXiv:1903.00510

\bibitem[{{Kaurov} {et~al}\mbox{.}(2019){Kaurov}, {Dai}, {Venumadhav},
  {Miralda-Escud{\'e}}, \& {Frye}}]{Kau2019}
{Kaurov} A.~A., {Dai} L., {Venumadhav} T., {Miralda-Escud{\'e}} J., {Frye} B.,
  2019, arXiv e-prints

\bibitem[{{Keeton} {et~al}\mbox{.}(1997){Keeton}, {Kochanek}, \&
  {Seljak}}]{Kee1997}
{Keeton} C.~R., {Kochanek} C.~S., {Seljak} U., 1997, \apj, 482, 604

\bibitem[{{Kelly} {et~al}\mbox{.}(2016){Kelly}, {Brammer}, {Selsing}, {Foley},
  {Hjorth}, {Rodney}, {Christensen}, {Strolger}, {Filippenko}, {Treu},
  {Steidel}, {Strom}, {Riess}, {Zitrin}, {Schmidt}, {Brada{\v c}}, {Jha},
  {Graham}, {McCully}, {Graur}, {Weiner}, {Silverman}, \& {Taddia}}]{Kel2016}
{Kelly} P.~L. {et~al.}, 2016, \apj, 831, 205

\bibitem[{{Kelly} {et~al}\mbox{.}(2018){Kelly}, {Diego}, {Rodney}, {Kaiser},
  {Broadhurst}, {Zitrin}, {Treu}, {P{\'e}rez-Gonz{\'a}lez}, {Morishita},
  {Jauzac}, {Selsing}, {Oguri}, {Pueyo}, {Ross}, {Filippenko}, {Smith},
  {Hjorth}, {Cenko}, {Wang}, {Howell}, {Richard}, {Frye}, {Jha}, {Foley},
  {Norman}, {Bradac}, {Zheng}, {Brammer}, {Benito}, {Cava}, {Christensen}, {de
  Mink}, {Graur}, {Grillo}, {Kawamata}, {Kneib}, {Matheson}, {McCully},
  {Nonino}, {P{\'e}rez-Fournon}, {Riess}, {Rosati}, {Schmidt}, {Sharon}, \&
  {Weiner}}]{Kel2018}
{Kelly} P.~L. {et~al.}, 2018, Nature Astronomy, 2, 334

\bibitem[{{Kelly} {et~al}\mbox{.}(2015){Kelly}, {Rodney}, {Treu}, {Foley},
  {Brammer}, {Schmidt}, {Zitrin}, {Sonnenfeld}, {Strolger}, {Graur},
  {Filippenko}, {Jha}, {Riess}, {Bradac}, {Weiner}, {Scolnic}, {Malkan}, {von
  der Linden}, {Trenti}, {Hjorth}, {Gavazzi}, {Fontana}, {Merten}, {McCully},
  {Jones}, {Postman}, {Dressler}, {Patel}, {Cenko}, {Graham}, \&
  {Tucker}}]{Kel2015}
{Kelly} P.~L. {et~al.}, 2015, Science, 347, 1123

\bibitem[{{Kochanek}(1991)}]{Koch1991}
{Kochanek} C.~S., 1991, \apj, 373, 354

\bibitem[{{Kolatt} \& {Bartelmann}(1998)}]{Kol1998}
{Kolatt} T.~S., {Bartelmann} M., 1998, \mnras, 296, 763

\bibitem[{{Kormann} {et~al}\mbox{.}(1994){Kormann}, {Schneider}, \&
  {Bartelmann}}]{Kor1994}
{Kormann} R., {Schneider} P., {Bartelmann} M., 1994, \aap, 284, 285

\bibitem[{{Levan} {et~al}\mbox{.}(2005){Levan}, {Nugent}, {Fruchter}, {Burud},
  {Branch}, {Rhoads}, {Castro-Tirado}, {Gorosabel}, {Castro Cer{\'o}n},
  {Thorsett}, {Kouveliotou}, {Golenetskii}, {Fynbo}, {Garnavich}, {Holland},
  {Hjorth}, {M{\o}ller}, {Pian}, {Tanvir}, {Ulanov}, {Wijers}, \&
  {Woosley}}]{Lev2005}
{Levan} A. {et~al.}, 2005, \apj, 624, 880

\bibitem[{{Li} {et~al}\mbox{.}(2011){Li}, {Leaman}, {Chornock}, {Filippenko},
  {Poznanski}, {Ganeshalingam}, {Wang}, {Modjaz}, {Jha}, {Foley}, \&
  {Smith}}]{Li2011}
{Li} W. {et~al.}, 2011, \mnras, 412, 1441

\bibitem[{{LSST Science Collaboration} {et~al}\mbox{.}(2009){LSST Science
  Collaboration}, {Abell}, {Allison}, {Anderson}, {Andrew}, {Angel}, {Armus},
  {Arnett}, {Asztalos}, {Axelrod}, {Bailey}, {Ballantyne}, {Bankert},
  {Barkhouse}, {Barr}, {Barrientos}, {Barth}, {Bartlett}, {Becker}, {Becla},
  {Beers}, {Bernstein}, {Biswas}, {Blanton}, {Bloom}, {Bochanski}, {Boeshaar},
  {Borne}, {Bradac}, {Brandt}, {Bridge}, {Brown}, {Brunner}, {Bullock},
  {Burgasser}, {Burge}, {Burke}, {Cargile}, {Chand rasekharan}, {Chartas},
  {Chesley}, {Chu}, {Cinabro}, {Claire}, {Claver}, {Clowe}, {Connolly}, {Cook},
  {Cooke}, {Cooray}, {Covey}, {Culliton}, {de Jong}, {de Vries}, {Debattista},
  {Delgado}, {Dell'Antonio}, {Dhital}, {Di Stefano}, {Dickinson}, {Dilday},
  {Djorgovski}, {Dobler}, {Donalek}, {Dubois-Felsmann}, {Durech},
  {Eliasdottir}, {Eracleous}, {Eyer}, {Falco}, {Fan}, {Fassnacht}, {Ferguson},
  {Fernandez}, {Fields}, {Finkbeiner}, {Figueroa}, {Fox}, {Francke}, {Frank},
  {Frieman}, {Fromenteau}, {Furqan}, {Galaz}, {Gal-Yam}, {Garnavich},
  {Gawiser}, {Geary}, {Gee}, {Gibson}, {Gilmore}, {Grace}, {Green}, {Gressler},
  {Grillmair}, {Habib}, {Haggerty}, {Hamuy}, {Harris}, {Hawley}, {Heavens},
  {Hebb}, {Henry}, {Hileman}, {Hilton}, {Hoadley}, {Holberg}, {Holman},
  {Howell}, {Infante}, {Ivezic}, {Jacoby}, {Jain}, {R}, {Jedicke}, {Jee},
  {Garrett Jernigan}, {Jha}, {Johnston}, {Jones}, {Juric}, {Kaasalainen},
  {Styliani}, {Kafka}, {Kahn}, {Kaib}, {Kalirai}, {Kantor}, {Kasliwal},
  {Keeton}, {Kessler}, {Knezevic}, {Kowalski}, {Krabbendam}, {Krughoff},
  {Kulkarni}, {Kuhlman}, {Lacy}, {Lepine}, {Liang}, {Lien}, {Lira}, {Long},
  {Lorenz}, {Lotz}, {Lupton}, {Lutz}, {Macri}, {Mahabal}, {Mandelbaum},
  {Marshall}, {May}, {McGehee}, {Meadows}, {Meert}, {Milani}, {Miller},
  {Miller}, {Mills}, {Minniti}, {Monet}, {Mukadam}, {Nakar}, {Neill}, {Newman},
  {Nikolaev}, {Nordby}, {O'Connor}, {Oguri}, {Oliver}, {Olivier}, {Olsen},
  {Olsen}, {Olszewski}, {Oluseyi}, {Padilla}, {Parker}, {Pepper}, {Peterson},
  {Petry}, {Pinto}, {Pizagno}, {Popescu}, {Prsa}, {Radcka}, {Raddick},
  {Rasmussen}, {Rau}, {Rho}, {Rhoads}, {Richards}, {Ridgway}, {Robertson},
  {Roskar}, {Saha}, {Sarajedini}, {Scannapieco}, {Schalk}, {Schindler},
  {Schmidt}, {Schmidt}, {Schneider}, {Schumacher}, {Scranton}, {Sebag},
  {Seppala}, {Shemmer}, {Simon}, {Sivertz}, {Smith}, {Allyn Smith}, {Smith},
  {Spitz}, {Stanford}, {Stassun}, {Strader}, {Strauss}, {Stubbs}, {Sweeney},
  {Szalay}, {Szkody}, {Takada}, {Thorman}, {Trilling}, {Trimble}, {Tyson}, {Van
  Berg}, {Vand en Berk}, {VanderPlas}, {Verde}, {Vrsnak}, {Walkowicz}, {Wand
  elt}, {Wang}, {Wang}, {Warner}, {Wechsler}, {West}, {Wiecha}, {Williams},
  {Willman}, {Wittman}, {Wolff}, {Wood-Vasey}, {Wozniak}, {Young}, {Zentner},
  \& {Zhan}}]{LSST2009}
{LSST Science Collaboration} {et~al.}, 2009, arXiv e-prints, arXiv:0912.0201

\bibitem[{{LSST Science Collaboration} {et~al}\mbox{.}(2017){LSST Science
  Collaboration}, {Marshall}, {Anguita}, {Bianco}, {Bellm}, {Brandt},
  {Clarkson}, {Connolly}, {Gawiser}, {Ivezic}, {Jones}, {Lochner}, {Lund},
  {Mahabal}, {Nidever}, {Olsen}, {Ridgway}, {Rhodes}, {Shemmer}, {Trilling},
  {Vivas}, {Walkowicz}, {Willman}, {Yoachim}, {Anderson}, {Antilogus}, {Angus},
  {Arcavi}, {Awan}, {Biswas}, {Bell}, {Bennett}, {Britt}, {Buzasi},
  {Casetti-Dinescu}, {Chomiuk}, {Claver}, {Cook}, {Davenport}, {Debattista},
  {Digel}, {Doctor}, {Firth}, {Foley}, {Fong}, {Galbany}, {Giampapa}, {Gizis},
  {Graham}, {Grillmair}, {Gris}, {Haiman}, {Hartigan}, {Hawley}, {Hlozek},
  {Jha}, {Johns-Krull}, {Kanbur}, {Kalogera}, {Kashyap}, {Kasliwal}, {Kessler},
  {Kim}, {Kurczynski}, {Lahav}, {Liu}, {Malz}, {Margutti}, {Matheson},
  {McEwen}, {McGehee}, {Meibom}, {Meyers}, {Monet}, {Neilsen}, {Newman},
  {O'Dowd}, {Peiris}, {Penny}, {Peters}, {Poleski}, {Ponder}, {Richards},
  {Rho}, {Rubin}, {Schmidt}, {Schuhmann}, {Shporer}, {Slater}, {Smith},
  {Soares-Santos}, {Stassun}, {Strader}, {Strauss}, {Street}, {Stubbs},
  {Sullivan}, {Szkody}, {Trimble}, {Tyson}, {de Val-Borro}, {Valenti},
  {Wagoner}, {Wood-Vasey}, \& {Zauderer}}]{LSST2017}
{LSST Science Collaboration} {et~al.}, 2017, arXiv e-prints

\bibitem[{{Madau} \& {Dickinson}(2014)}]{Mad2014}
{Madau} P., {Dickinson} M., 2014, \araa, 52, 415

\bibitem[{{Maoz} \& {Mannucci}(2012)}]{Mao2012}
{Maoz} D., {Mannucci} F., 2012, PASA, 29, 447

\bibitem[{{Melinder} {et~al}\mbox{.}(2012){Melinder}, {Dahlen}, {Menc{\'{\i}}a
  Trinchant}, {{\"O}stlin}, {Mattila}, {Sollerman}, {Fransson}, {Hayes},
  {Kankare}, \& {Nasoudi-Shoar}}]{Mel2012}
{Melinder} J. {et~al.}, 2012, \aap, 545, A96

\bibitem[{{Mitchell} {et~al}\mbox{.}(2005){Mitchell}, {Keeton}, {Frieman}, \&
  {Sheth}}]{Mitch2005}
{Mitchell} J.~L., {Keeton} C.~R., {Frieman} J.~A., {Sheth} R.~K., 2005, \apj,
  622, 81

\bibitem[{{Montero-Dorta} {et~al}\mbox{.}(2017){Montero-Dorta}, {Bolton}, \&
  {Shu}}]{Mont2017}
{Montero-Dorta} A.~D., {Bolton} A.~S., {Shu} Y., 2017, \mnras, 468, 47

\bibitem[{{More} {et~al}\mbox{.}(2017){More}, {Suyu}, {Oguri}, {More}, \&
  {Lee}}]{2017Mor}
{More} A., {Suyu} S.~H., {Oguri} M., {More} S., {Lee} C.-H., 2017, \apj, 835,
  L25

\bibitem[{{Nugent} {et~al}\mbox{.}(2002){Nugent}, {Kim}, \&
  {Perlmutter}}]{Nug2002}
{Nugent} P., {Kim} A., {Perlmutter} S., 2002, \pasp, 114, 803

\bibitem[{{Oguri}(2010)}]{Ogu2010}
{Oguri} M., 2010, \pasj, 62, 1017

\bibitem[{{Oguri} {et~al}\mbox{.}(2008){Oguri}, {Inada}, {Strauss}, {Kochanek},
  {Richards}, {Schneider}, {Becker}, {Fukugita}, {Gregg}, {Hall}, {Hennawi},
  {Johnston}, {Kayo}, {Keeton}, {Pindor}, {Shin}, {Turner}, {White}, {York},
  {Anderson}, {Bahcall}, {Brunner}, {Burles}, {Castander}, {Chiu},
  {Clocchiatti}, {Eisenstein}, {Frieman}, {Kawano}, {Lupton}, {Morokuma},
  {Rix}, {Scranton}, \& {Sheldon}}]{Ogu2008}
{Oguri} M. {et~al.}, 2008, \aj, 135, 512

\bibitem[{{Oguri} \& {Kawano}(2003)}]{Ogu2003}
{Oguri} M., {Kawano} Y., 2003, \mnras, 338, L25

\bibitem[{{Oguri} \& {Marshall}(2010)}]{Ogu2010a}
{Oguri} M., {Marshall} P.~J., 2010, \mnras, 405, 2579

\bibitem[{{Pain} {et~al}\mbox{.}(2002){Pain}, {Fabbro}, {Sullivan}, {Ellis},
  {Aldering}, {Astier}, {Deustua}, {Fruchter}, {Goldhaber}, {Goobar}, {Groom},
  {Hardin}, {Hook}, {Howell}, {Irwin}, {Kim}, {Kim}, {Knop}, {Lee}, {Lidman},
  {McMahon}, {Nugent}, {Panagia}, {Pennypacker}, {Perlmutter}, {Ruiz-Lapuente},
  {Schahmaneche}, {Schaefer}, \& {Walton}}]{Pai2002}
{Pain} R. {et~al.}, 2002, \apj, 577, 120

\bibitem[{{Perrett} {et~al}\mbox{.}(2012){Perrett}, {Sullivan}, {Conley},
  {Gonz{\'a}lez-Gait{\'a}n}, {Carlberg}, {Fouchez}, {Ripoche}, {Neill},
  {Astier}, {Balam}, {Balland}, {Basa}, {Guy}, {Hardin}, {Hook}, {Howell},
  {Pain}, {Palanque-Delabrouille}, {Pritchet}, {Regnault}, {Rich},
  {Ruhlmann-Kleider}, {Baumont}, {Lidman}, {Perlmutter}, \& {Walker}}]{Per2012}
{Perrett} K. {et~al.}, 2012, \aj, 144, 59

\bibitem[{{Pierel} \& {Rodney}(2019)}]{Pie2019}
{Pierel} J.~R., {Rodney} S.~A., 2019, arXiv e-prints, arXiv:1902.01260

\bibitem[{{Quimby} {et~al}\mbox{.}(2014){Quimby}, {Oguri}, {More}, {More},
  {Moriya}, {Werner}, {Tanaka}, {Folatelli}, {Bersten}, {Maeda}, \&
  {Nomoto}}]{Qui2014}
{Quimby} R.~M. {et~al.}, 2014, Science, 344, 396

\bibitem[{{Refsdal}(1964)}]{Ref1964}
{Refsdal} S., 1964, \mnras, 128, 307

\bibitem[{{Richardson} {et~al}\mbox{.}(2014{\natexlab{a}}){Richardson},
  {Jenkins}, {Wright}, \& {Maddox}}]{Rich2014}
{Richardson} D., {Jenkins}, III R.~L., {Wright} J., {Maddox} L.,
  2014{\natexlab{a}}, \aj, 147, 118

\bibitem[{{Richardson} {et~al}\mbox{.}(2014{\natexlab{b}}){Richardson},
  {Jenkins}, {Wright}, \& {Maddox}}]{2014AJ....147..118R}
{Richardson} D., {Jenkins}, III R.~L., {Wright} J., {Maddox} L.,
  2014{\natexlab{b}}, \aj, 147, 118

\bibitem[{{Rodney} {et~al}\mbox{.}(2018){Rodney}, {Balestra}, {Bradac},
  {Brammer}, {Broadhurst}, {Caminha}, {Chiriv{\i}}, {Diego}, {Filippenko},
  {Foley}, {Graur}, {Grillo}, {Hemmati}, {Hjorth}, {Hoag}, {Jauzac}, {Jha},
  {Kawamata}, {Kelly}, {McCully}, {Mobasher}, {Molino}, {Oguri}, {Richard},
  {Riess}, {Rosati}, {Schmidt}, {Selsing}, {Sharon}, {Strolger}, {Suyu},
  {Treu}, {Weiner}, {Williams}, \& {Zitrin}}]{Rod2018}
{Rodney} S.~A. {et~al.}, 2018, Nature Astronomy, 2, 324

\bibitem[{{Rodney} {et~al}\mbox{.}(2015){Rodney}, {Patel}, {Scolnic}, {Foley},
  {Molino}, {Brammer}, {Jauzac}, {Brada{\v c}}, {Broadhurst}, {Coe}, {Diego},
  {Graur}, {Hjorth}, {Hoag}, {Jha}, {Johnson}, {Kelly}, {Lam}, {McCully},
  {Medezinski}, {Meneghetti}, {Merten}, {Richard}, {Riess}, {Sharon},
  {Strolger}, {Treu}, {Wang}, {Williams}, \& {Zitrin}}]{Rod2015}
{Rodney} S.~A. {et~al.}, 2015, \apj, 811, 70

\bibitem[{{Rodney} {et~al}\mbox{.}(2014){Rodney}, {Riess}, {Strolger},
  {Dahlen}, {Graur}, {Casertano}, {Dickinson}, {Ferguson}, {Garnavich},
  {Hayden}, {Jha}, {Jones}, {Kirshner}, {Koekemoer}, {McCully}, {Mobasher},
  {Patel}, {Weiner}, {Cenko}, {Clubb}, {Cooper}, {Filippenko}, {Frederiksen},
  {Hjorth}, {Leibundgut}, {Matheson}, {Nayyeri}, {Penner}, {Trump},
  {Silverman}, {U}, {Azalee Bostroem}, {Challis}, {Rajan}, {Wolff}, {Faber},
  {Grogin}, \& {Kocevski}}]{Rod2014}
{Rodney} S.~A. {et~al.}, 2014, \aj, 148, 13

\bibitem[{{Rodney} {et~al}\mbox{.}(2016){Rodney}, {Strolger}, {Kelly},
  {Brada{\v c}}, {Brammer}, {Filippenko}, {Foley}, {Graur}, {Hjorth}, {Jha},
  {McCully}, {Molino}, {Riess}, {Schmidt}, {Selsing}, {Sharon}, {Treu},
  {Weiner}, \& {Zitrin}}]{Rod2016}
{Rodney} S.~A. {et~al.}, 2016, \apj, 820, 50

\bibitem[{{Rodney} \& {Tonry}(2010)}]{Rod2010}
{Rodney} S.~A., {Tonry} J.~L., 2010, \apj, 723, 47

\bibitem[{{Schneider} \& {Sluse}(2014)}]{Sch2014}
{Schneider} P., {Sluse} D., 2014, \aap, 564, A103

\bibitem[{{Smith}(2014)}]{2014ARA&A..52..487S}
{Smith} N., 2014, \araa, 52, 487

\bibitem[{{Taddia} {et~al}\mbox{.}(2013){Taddia}, {Stritzinger}, {Sollerman},
  {Phillips}, {Anderson}, {Boldt}, {Campillay}, {Castell{\'o}n}, {Contreras},
  {Folatelli}, {Hamuy}, {Heinrich-Josties}, {Krzeminski}, {Morrell}, {Burns},
  {Freedman}, {Madore}, {Persson}, \& {Suntzeff}}]{2013A&A...555A..10T}
{Taddia} F. {et~al.}, 2013, \aap, 555, A10

\bibitem[{{Th{\"o}ne} {et~al}\mbox{.}(2017){Th{\"o}ne}, {de Ugarte Postigo},
  {Leloudas}, {Gall}, {Cano}, {Maeda}, {Schulze}, {Campana}, {Wiersema},
  {Groh}, {de la Rosa}, {Bauer}, {Malesani}, {Maund}, {Morrell}, \&
  {Beletsky}}]{2017A&A...599A.129T}
{Th{\"o}ne} C.~C. {et~al.}, 2017, \aap, 599, A129

\bibitem[{{Tonry} {et~al}\mbox{.}(2003){Tonry}, {Schmidt}, {Barris}, {Candia},
  {Challis}, {Clocchiatti}, {Coil}, {Filippenko}, {Garnavich}, {Hogan},
  {Holland}, {Jha}, {Kirshner}, {Krisciunas}, {Leibundgut}, {Li}, {Matheson},
  {Phillips}, {Riess}, {Schommer}, {Smith}, {Sollerman}, {Spyromilio},
  {Stubbs}, \& {Suntzeff}}]{Ton2003}
{Tonry} J.~L. {et~al.}, 2003, \apj, 594, 1

\bibitem[{{Vega-Ferrero} {et~al}\mbox{.}(2018){Vega-Ferrero}, {Diego},
  {Miranda}, \& {Bernstein}}]{Vega2018}
{Vega-Ferrero} J., {Diego} J.~M., {Miranda} V., {Bernstein} G.~M., 2018, \apjl,
  853, L31

\bibitem[{{Witt} \& {Mao}(1997)}]{Witt1997}
{Witt} H.~J., {Mao} S., 1997, \mnras, 291, 211

\bibitem[{{Yahalomi} {et~al}\mbox{.}(2017){Yahalomi}, {Schechter}, \&
  {Wambsganss}}]{Yah2017}
{Yahalomi} D.~A., {Schechter} P.~L., {Wambsganss} J., 2017, arXiv e-prints,
  arXiv:1711.07919

\end{thebibliography}

\appendix
\onecolumn

\begin{table*}
\section{Parameters of fitting function}
\begin{tabular}{lccrrrrr}
Method & SN Type & Bandpass & $X_{0}$ & $X_{1}$ & $X_{2}$ & $X_{3}$  & $X_{4}$ \\
\hline
magnification & Ia & $g$ &  0.72036 &  0.24009 & -0.03900 &  0.00477 &  0.00104  \\
magnification & Ia & $r$ & 1.33862 &  0.19880 & -0.04735 &  0.00450 & -0.00012  \\
magnification & Ia & $i$ & 1.83059 &  0.21667 & -0.09311 &  0.00355 &  0.00255  \\
magnification & Ia & $z$ & 2.07615 &  0.25577 & -0.10111 &  0.00216 &  0.00285  \\
magnification & Ia & $y$ & 2.23439 &  0.29536 & -0.10980 &  0.00041 &  0.00343  \\
\hline
multiple images & Ia & $g$ &  0.48173 &  0.61185 & -0.06134 &  0.00563 & -0.00097  \\
multiple images & Ia & $r$ & 1.07923 &  0.67684 & -0.07377 &  0.00196 & -0.00009  \\
multiple images & Ia & $i$ & 1.33432 &  0.80388 & -0.09728 & -0.01051 &  0.00351  \\
multiple images & Ia & $z$ & 1.38904 &  0.88393 & -0.08518 & -0.01878 &  0.00392  \\
multiple images & Ia & $y$ & 1.38537 &  0.94450 & -0.07931 & -0.02503 &  0.00464  \\
\hline
hybrid & Ia & $g$ & 0.88408 &  0.37177 & -0.00971 &  0.00149 & -0.00033  \\
hybrid & Ia & $r$ & 1.47731 &  0.36007 &  0.00447 &  0.00321 & -0.00195  \\
hybrid & Ia & $i$ & 1.89609 &  0.35927 & -0.02158 &  0.00505 & -0.00031  \\
hybrid & Ia & $z$ & 2.10720 &  0.37109 & -0.03153 &  0.00530 &  0.00018  \\
hybrid & Ia & $y$ & 2.24285 &  0.38345 & -0.04630 &  0.00507 &  0.00118  \\
\hline
\hline
magnification & IIP & $g$ & -0.09803 &  0.30353 & -0.03839 & -0.00170 &  0.00210  \\
magnification & IIP & $r$ &  0.51297 &  0.18132 & -0.03835 &  0.00365 & -0.00023  \\
magnification & IIP & $i$ & 1.05567 &  0.17664 & -0.07500 &  0.00283 &  0.00206  \\
magnification & IIP & $z$ &  1.37635 &  0.22296 & -0.08338 &  0.00011 &  0.00265  \\
magnification & IIP & $y$ &  1.60361 &  0.26657 & -0.09178 & -0.00201 &  0.00324  \\
\hline
multiple image & IIP & $g$ & -0.73742 &  0.74209 & -0.02966 & -0.00791 &  0.00130  \\
multiple image & IIP & $r$ & 0.05035 &  0.79421 & -0.05724 & -0.01589 &  0.00452  \\
multiple image & IIP & $i$ & 0.39685 &  0.85468 & -0.06020 & -0.01743 &  0.00460  \\
multiple image & IIP & $z$ & 0.53709 &  0.87378 & -0.04444 & -0.01662 &  0.00342  \\
multiple image & IIP & $y$ & 0.62664 &  0.86039 & -0.03253 & -0.01300 &  0.00225  \\
\hline
hybrid & IIP & $g$ &  -0.03277 &  0.39937 &  0.00269 & -0.00081 &  0.00066  \\
hybrid & IIP & $r$ &  0.61623 &  0.33920 &  0.02793 &  0.00642 & -0.00203  \\
hybrid & IIP & $i$ &  1.10586 &  0.32201 &  0.00987 &  0.00839 & -0.00064  \\
hybrid & IIP & $z$ &  1.39204 &  0.32922 & -0.00632 &  0.00811 &  0.00059  \\
hybrid & IIP & $y$ &  1.60046 &  0.33684 & -0.02764 &  0.00781 &  0.00209  \\
\hline
\hline
magnification & IIL & $g$ & -0.40183 &  0.23634 & -0.01517 &  0.00582 & -0.00057  \\
magnification & IIL & $r$ & 0.25718 &  0.15857 & -0.03946 &  0.00435 & -0.00008  \\
magnification & IIL & i & 0.86956 &  0.17352 & -0.07717 &  0.00352 &  0.00206  \\
magnification & IIL & z & 1.22347 &  0.22128 & -0.08761 &  0.00140 &  0.00264  \\
magnification & IIL & y & 1.46877 &  0.26402 & -0.09947 & -0.00036 &  0.00340  \\
\hline
multiple image & IIL & g &-0.74372 &  0.65666 & -0.05661 & -0.00358 &  0.00195  \\
multiple image & IIL & r & 0.07288 &  0.66921 & -0.06623 & -0.00301 &  0.00203  \\
multiple image & IIL & i & 0.47375 &  0.72861 & -0.07291 & -0.00444 &  0.00220  \\
multiple image & IIL & z & 0.65237 &  0.77378 & -0.06863 & -0.00711 &  0.00228  \\
multiple image & IIL & y & 0.78459 &  0.79863 & -0.05759 & -0.00825 &  0.00117  \\
\hline
hybrid & IIL & g & -0.26604 &  0.36248 &  0.01242 &  0.00311 & -0.00151  \\
hybrid & IIL & r & 0.41663 &  0.35667 &  0.01112 &  0.00232 & -0.00097  \\
hybrid & IIL & i & 0.95985 &  0.34048 &  0.00086 &  0.00553 & -0.00076  \\
hybrid & IIL & z & 1.26577 &  0.34858 & -0.01301 &  0.00577 &  0.00010  \\
hybrid & IIL & y & 1.48178 &  0.35631 & -0.03247 &  0.00597 &  0.00141  \\

\end{tabular}
\caption{Best fit parameters of a polynomial function approximating the computed 
rates of detecting strongly lensed supernovae, for different detection methods and bandpass filters. 
The fitting function is given by eq.~(\ref{detection_model}) and the rates 
are in units of yr$^{-1}$~$(4\pi)^{-1}$ (yields from a full-sky survey per year).}
\label{appro}
\end{table*}

\begin{table*}
\begin{tabular}{lccrrrrr}
Method & SN Type & Bandpass & $X_{0}$ & $X_{1}$ & $X_{2}$ & $X_{3}$  & $X_{4}$ \\
\hline
magnification & Ibc & $g$ & 0.14709 &  0.31974 &  0.00689 &  0.00305 & -0.00104  \\
magnification & Ibc & $r$ & 0.49563 &  0.20613 & -0.02187 &  0.00285 & -0.00080  \\
magnification & Ibc & $i$ & 1.02766 &  0.17760 & -0.07830 &  0.00321 &  0.00219  \\
magnification & Ibc & $z$ & 1.37074 &  0.22244 & -0.08762 &  0.00128 &  0.00264  \\
magnification & Ibc & $y$ & 1.61716 &  0.26280 & -0.09812 & -0.00027 &  0.00329  \\
\hline
multiple images & Ibc & $g$ &  -0.67826 &  0.63296 & -0.01858 & -0.00064 & -0.00025  \\
multiple images & Ibc & $r$ & 0.16955 &  0.68416 & -0.07132 & -0.00334 &  0.00255  \\
multiple images & Ibc & $i$ & 0.55571 &  0.74754 & -0.07414 & -0.00661 &  0.00290  \\
multiple images & Ibc & $z$ & 0.73853 &  0.75590 & -0.05557 & -0.00484 &  0.00125  \\
multiple images & Ibc & $y$ & 0.85114 &  0.77360 & -0.05053 & -0.00659 &  0.00142  \\
 \hline
hybrid & Ibc & $g$ & 0.19634 &  0.35951 &  0.01349 &  0.00263 & -0.00104  \\
hybrid & Ibc & $r$ & 0.62451 &  0.34332 &  0.01745 &  0.00303 & -0.00160  \\
hybrid & Ibc & $i$ & 1.09885 &  0.32867 & -0.00133 &  0.00634 & -0.00050  \\
hybrid & Ibc & $z$ & 1.40103 &  0.33547 & -0.01531 &  0.00681 &  0.00035  \\
hybrid & Ibc & $y$ & 1.62488 &  0.34237 & -0.03615 &  0.00692 &  0.00172  \\
\hline
\hline
magnification & IIn & $g$ &  1.90814 &  0.45380 & -0.03934 & -0.00427 &  0.00045  \\
magnification & IIn & $r$ & 1.91461 &  0.38611 & -0.05516 & -0.00382 &  0.00061  \\
magnification & IIn & $i$ & 1.95587 &  0.31645 & -0.10087 & -0.00210 &  0.00325  \\
magnification & IIn & $z$ & 1.98209 &  0.31814 & -0.10942 & -0.00231 &  0.00377  \\
magnification & IIn & $y$ & 2.01341 &  0.33552 & -0.10991 & -0.00248 &  0.00357  \\
\hline
multiple images & IIn & $g$ &  1.11411 &  0.72917 & -0.04473 & -0.00435 &  0.00007  \\
multiple images & IIn & $r$ &  1.18910 &  0.75405 & -0.05117 & -0.00637 &  0.00056  \\
multiple images & IIn & $i$ &  1.19611 &  0.78006 & -0.05496 & -0.00828 &  0.00096  \\
multiple images & IIn & $z$ &  1.17156 &  0.79957 & -0.05529 & -0.00965 &  0.00115  \\
multiple images & IIn & $y$ &  1.13596 &  0.81564 & -0.05541 & -0.01082 &  0.00136  \\
\hline
hybrid & IIn & $g$ & 1.91431 &  0.46530 & -0.03247 & -0.00292 &  0.00050  \\
hybrid & IIn & $r$ & 1.94418 &  0.42431 & -0.03745 & -0.00056 &  0.00074  \\
hybrid & IIn & $i$ & 1.96332 &  0.37749 & -0.05314 &  0.00379 &  0.00213  \\
hybrid & IIn & $z$ & 1.98247 &  0.37635 & -0.05987 &  0.00410 &  0.00260  \\
hybrid & IIn & $y$ & 2.00122 &  0.38171 & -0.06689 &  0.00395 &  0.00299  \\

\end{tabular}
\caption{Continuation of Table~\ref{appro}.}
\label{appro1}
\end{table*}

\end{document}